\documentclass{aastex}
\usepackage{emulateapj5}

\usepackage{textcomp}
\usepackage{amssymb,amsmath}

\newcommand{\FeXXV}{\hbox{Fe {\sc xxv}}}
\newcommand{\FeXXVI}{\hbox{Fe {\sc xxvi}}}

\newcommand{\simgt}{\lower 2pt \hbox{$\, \buildrel {\scriptstyle >}\over {\scriptstyle\sim}\,$}}
\newcommand{\simlt}{\lower 2pt \hbox{$\, \buildrel {\scriptstyle <}\over {\scriptstyle\sim}\,$}}

\newcommand{\apm}{APM~08279+5255}

\newcommand{\chandra}{{\emph{Chandra}}}

\newcommand{\xmm}{\emph{XMM-Newton}}

\newcommand{\lnh}{\hbox{{\rm log}~$N_{\rm H}$}}

\slugcomment{Received 2009 May 12, Accepted 2009 September 28}
\shorttitle{Confirmation of Near-Relativistic Outflow}
\shortauthors{CHARTAS ET AL.}

\begin{document}

\def\sarc{$^{\prime\prime}\!\!.$}
\def\arcsec{$^{\prime\prime}$}
\def\beginrefer{\section*{References}%
\begin{quotation}\mbox{}\par}
\def\refer#1\par{{\setlength{\parindent}{-\leftmargin}\indent#1\par}}
\def\endrefer{\end{quotation}}




\title{Confirmation of and Variable Energy Injection by a Near-Relativistic Outflow in \apm}

\author{G. Chartas,\altaffilmark{1,2} C. Saez,\altaffilmark{2} W. N. Brandt,\altaffilmark{2} M. Giustini,\altaffilmark{2,3,4} and G. P. Garmire\altaffilmark{2} } 


\altaffiltext{1}{Department of Physics and Astronomy, College of Charleston, Charleston, SC, 29424, USA, chartasg@cofc.edu}

\altaffiltext{2}{Department of Astronomy \& Astrophysics, Pennsylvania State University,
University Park, PA 16802, chartas@astro.psu.edu, saez@astro.psu.edu, niel@astro.psu.edu, giustini@astro.psu.edu, garmire@astro.psu.edu}

\altaffiltext{3}{Dipartimento di Astronomia, Universita degli Studi di Bologna, via Ranzani 1, 40127 Bologna, Italy, 
margharita.giustini2@iasfbo.inaf.it}

\altaffiltext{4}{INAF/Istituto di Astrofisica Spaziale e Fisica Cosmica di Bologna, via Gobetti 101, 40129 Bologna, Italy, margherita.giustini@unibo.it}


\begin{abstract}

We present results from multi-epoch spectral analysis of {\sl XMM-Newton} 
and {\sl Chandra} observations of the broad absorption line (BAL) quasar \apm.
Our analysis shows significant X-ray BALs in all epochs
with rest-frame energies lying in the range of $\sim$ 6.7--18~keV.
The X-ray BALs and 0.2--10~keV continuum show significant variability
on timescales as short as 3.3~days (proper time) 
implying a source size-scale of $\sim$ 10~$r_{\rm g}$, where  $r_{\rm g}$ is the gravitational radius.
We find a large gradient in the outflow velocity of the X-ray absorbers with projected 
outflow velocities of up to 0.76~$c$.
The maximum outflow velocity constrains the angle between the wind velocity and 
our line of sight to be less than $\sim$ 22$^{\circ}$.
Based on our spectral analysis we identify the following components of the outflow:
(a) Highly ionized X-ray absorbing material with an ionization parameter in the range 
of $2.9 \simlt \log\xi \simlt 3.9$ (the units of $\xi$ are erg~cm~s$^{-1}$) and a column density of  $\log{N_{\rm H}} \sim 23$ (the units of  $N_{\rm H}$ are cm$^{-2}$)
outflowing at velocities of up to 0.76~$c$. (b) Low-ionization X-ray absorbing gas with 
$\log{N_{\rm H}} \sim 22.8$.
We find a possible trend between the X-ray photon index and 
the maximum outflow velocity of the ionized 
absorber in the sense that flatter 
spectra appear to result in lower outflow velocities.
Based on our spectral analysis of observations of \apm\
over a period of 1.2 years (proper time) we estimate the mass-outflow rate 
and efficiency of the outflow to have varied 
between $16_{-8}^{+12}$ $M_{\odot}$~yr$^{-1}$ and $64_{-40}^{+66}$ $M_{\odot}$~yr$^{-1}$ and $0.18_{-0.11}^{+0.15}$  to $1.7_{-1.2}^{+1.9}$, respectively. 
Assuming that the outflow properties of  \apm\ are 
a common property of most quasars at similar redshifts, our results then imply that quasar winds
are massive and energetic enough
to influence significantly the formation of the host galaxy,
provide significant metal enrichment to the 
interstellar medium (ISM) and intergalactic medium (IGM), and
are a viable mechanism for feedback at redshifts
near the peak in the number density of galaxy mergers.

\end{abstract}

\keywords{galaxies: active --- quasars: absorption lines --- quasars: 
individual~(APM~08279+5225) --- X-rays: galaxies --- gravitational lensing}

\section{INTRODUCTION}

A number of physical mechanisms of interactions of AGNs with their environments
have been proposed to explain feedback in galaxies and clusters of galaxies.
These feedback mechanisms include radio jets, AGN winds, and AGN radiative 
heating. 
It may be the case that several of these mechanisms operate simultaneously and their relative 
contribution to the feedback process varies depending in part on the 
size of the super-massive black hole (SMBH), and the combined gravitational potential of the SMBH and host galaxy.
The gas and dust content of the host galaxy, the redshift 
and age of the host, and the inflow and outflow properties of the central super-massive black hole may be important as well.

The radio-jet feedback mechanism was proposed after the discovery of
significant radio cavities and shock fronts in several clusters of galaxies. 
The cavity sizes are large enough to imply that the amount of energy injected in 
the intracluster medium (ICM), as inferred from the mechanical work $pdV$ required to displace the gas in the radio cavities,
is sufficient to balance radiative losses, suppress cooling flows, and halt star formation (e.g., Fabian et al. 2009, and references therein).
The radio jets are thought to be driven by either accretion 
onto the black hole (e.g., Blandford \& Payne 1982) or the black-hole spin (e.g., Blandford \& Znajek 1977).
The details of the process by which radio jets deposit energy 
and distribute it uniformly throughout the ICM is currently not well understood. 

Another possibly important mode that may contribute to feedback is AGN radiation.
Radiative heating (e.g., Ciotti \& Ostriker 2007)
has been proposed as an alternative to the radio-jet mechanism to explain the suppression of cooling flows in isolated elliptical galaxies and possibly in clusters of galaxies. 
 
 
In field galaxies, especially ones in the redshift
range of $z \approx 1-3 $ where the number density of galaxy mergers is
thought to peak (e.g., Di Matteo et al. 2005), quasar winds 
are thought to be one of the major contributors to feedback.
Winds can be driven by radiation pressure, magneto-centrifugal forces, and
thermal pressure or a combination of these processes. The potential importance of quasar outflows has been explicitly demonstrated
in theoretical models of structure formation and galaxy 
mergers that incorporate the effects of quasar outflows 
(e.g., Silk \& Rees 1998; Scannapieco \& Oh 2004; Granato et al. 2004; 
Springel, Di Matteo, \& Hernquist 2005; Hopkins et al. 2005, 2006). 
Observational evidence to support these theories of quasar outflows was 
provided by the discovery of near-relativistic X-ray absorbing 
winds in several broad absorption line (BAL)
quasars (e.g., Chartas et al. 2002, 2003) and later in several non-BAL quasars 
(e.g., Reeves et al. 2003; Pounds et al. 2003, 2006; Braito et al. 2007).

The relative contribution of each of these proposed feedback mechanisms will
vary in part depending on the following conditions:

(a) The importance of the radio-jet mode 
will depend on the radio loudness and the duty cycle of the central AGN.
The presence of a radio jet in an AGN is likely to depend
on the available supply of infalling material to feed the black hole and/or the 
spin of the black hole. We note that several observations imply that the 
fraction of radio-loud objects varies with redshift and luminosity
(e.g., Peacock et al. 1986; Schneider et al. 1992; La Franca et al. 1994; Jiang et al. 2007).
Specifically it is found that the radio-loud fraction decreases with redshift and increases with luminosity. The radio jet mode therefore may be more important at later times 
and in clusters of galaxies and in nearby massive ellipticals that 
have a larger probability of containing a radio-loud object compared 
to galaxies at $ z {\simgt}  1$ which are more likely to contain a radio-quiet AGN.


(b) The feedback process is also expected to depend on the mass of the central black hole and the 
gravitational potential of the host galaxy.
In order for a feedback mechanism to be effective it must be able to
drive gas out of the host with a velocity that is larger than 
the escape velocity determined by the gravitational potential of the system.
The mass outflow rates and outflow efficiencies of AGN winds will also depend 
strongly on the mass and luminosity of the SMBH.
Analyses of the kinematic properties of UV absorbers 
in quasars indicate a strong increase in outflow velocity of UV absorbers 
with quasar UV luminosity (e.g., Laor \& Brandt 2002; Ganguly \& Brotherton 2008).
Luminous quasars are found to contain winds with velocities of the UV outflowing material of 
up to 60,000~km~s$^{-1}$ whereas Seyfert galaxies typically have 
winds with significantly lower velocities of up to $\sim$ 2,000 km~s$^{-1}$.

(c) Hydrodynamic simulations of AGN atmospheres driven by radiation pressure indicate that 
the strengths of the outflows originating from AGN accretion disks depend on the 
Eddington ratio ($L_{\rm Bol}/L_{\rm Edd}$) with more massive and faster winds produced at larger Eddington ratios
(e.g., Proga, Stone \& Drew 1988; Proga, Stone \& Kallman 2000; Proga \& Kallman 2004).
In systems with a limited supply of gas to fuel the AGN 
it is expected that AGN winds will be relatively weak.

(d) In the case of a galaxy merger the 
degree to which each mechanism contributes to the overall feedback
process may change with time since the merger event occurred.

Observations in X-rays of the BAL quasar APM~08279+5255, the mini-BAL quasar
PG~1115+080, and perhaps the low-ionization BAL quasar H~1413-117 have strongly suggested
the presence of near-relativistic outflows of X-ray absorbing
material in these objects (Chartas et al. 2002, 2003, 2007a, 2007b).
Observations of these quasars in the optical and UV indicate
outflow velocities and column densities that are more than an order of magnitude 
lower than those inferred from the X-ray spectra.
These differences in outflow velocities and absorbing column densities
suggest that X-ray BALs probe a highly ionized, massive and high-velocity 
component of the wind that is largely distinct from the absorbers detected in the optical 
and UV wavebands. 
Most of the mass and energy of quasar winds is likely
carried by the outflowing X-ray absorbers, and
observations in the X-ray band are therefore crucial 
for improving our understanding of the contribution of quasar winds 
to feedback in galaxies.

This paper will focus on determining the significance of the wind in
the $z = 3.91$ gravitationally lensed BAL quasar \apm\ in the feedback process during an epoch near the peak of the number density of mergers. Specifically, we present results from
analyses of several \xmm\ and \chandra\ observations of \apm\ in order to
determine the outflow velocity and its variability, 
obtain insights on the acceleration mechanism,
constrain the geometry of the wind, and estimate the mass outflow rate and
the outflow efficiency.

Throughout this paper we adopt a $\Lambda$-dominated cosmology with $H_{0}$ = 70~km~s$^{-1}$~Mpc$^{-1}$, 
$\Omega_{\rm \Lambda}$ = 0.7, and  $\Omega_{\rm M}$ = 0.3.

\section{OBSERVATIONS AND DATA ANALYSIS}

\subsection{Observations and Data Reduction}

\apm\ was observed with \xmm\ (Jansen et al. 2001) on 2002 April 28, 2007 October 6, and
2007 October 22, respectively.
It was also observed with the Advanced CCD Imaging 
Spectrometer (ACIS; Garmire et al. 2003) on board the {\it Chandra X-ray Observatory} 
(hereafter \chandra) on 2002 February 24 and 2008 January 14.
 
 The spectral analysis of the 2002 \chandra\ observation of \apm\ has been presented in
Chartas et al. (2002), and the spectral analysis of the 
2002 \xmm\ observation of \apm\ in Hasinger et al. (2002).
Because of recent significant improvements in the calibration of the
instruments on board \chandra\ and \xmm\ since the publication of the 
\apm\ results, we have re-analyzed all observations.
Updates on the calibration of \chandra\ and \xmm\ are reported 
on the \chandra\ X-ray Center (CXC) and \xmm\ Science Operations Centre (SOC) 
World Wide Web (WWW) sites, respectively.\footnote{The WWW sites listing the updates are located at \url{http://asc.harvard.edu/ciao/releasenotes/history.html } 
and \url{http://xmm2.esac.esa.int/external/xmm\_sw\_cal/calib/},  respectively.}

We analyzed the \xmm\ data for \apm\ with the standard analysis software SAS version 8.0.0 
provided by the \xmm\ SOC.  
The \chandra\ observations of \apm\ were analyzed using the standard software 
CIAO 4.1 provided by the CXC. 
A log of the observations that includes 
observation dates, observed count rates, 
total exposure times,  and observational identification numbers is presented in Table 1. 


We note that the count rate and total number of counts 
($\sim$ 40,918 counts from all \xmm\ observations)  for \apm\ are the highest of any BAL 
quasar X-ray spectrum observed to date.
This can be compared to the typical total number of 0.5--8~keV counts 
detected in BAL quasars obtained from the Sloan digital sky survey data release 5
that ranges between a few to $\sim$ 200 counts (Gibson et al. 2009).

For the reduction of the \xmm\ observations we filtered the 
pn (Str{\" u}der et al. 2001) and MOS (Turner et al. 2001)
data by selecting events corresponding to instrument \verb+PATTERNS+
in the 0--4 (single and double pixel events) and 0--12 (up to
quadruple pixel events) ranges, respectively.
Several moderate-amplitude background flares were present
during the \xmm\ observations of \apm.
The pn and MOS data were filtered to exclude times when the full-field-of-view count rates
exceeded 20~cnts~s$^{-1}$ and 4~cnts~s$^{-1}$, respectively.   
The extracted spectra from the pn and MOS 
were grouped to obtain a minimum of 100 counts in each energy bin,
allowing use of $ \chi^{2}$ statistics.
Background spectra for the pn and MOS detectors were extracted
from source-free regions near \apm.

The pn and MOS spectra were fit with a variety of models employing 
\verb+XSPEC+ version 12.5 (Arnaud 1996).  The energy ranges used for fitting the pn and MOS 
data were 0.2--10~keV and 0.4--10~keV, respectively.
We performed spectral fits to the pn spectra alone
and to the pn and MOS data simultaneously.
Both approaches resulted in values for the fitted parameters that were 
consistent within the errors, however, the fits to the higher quality pn data alone provided 
higher quality fits as indicated by the reduced $\chi^{2}$ values of these fits.
We therefore consider the results from the
fits to the pn data alone more reliable especially for 
characterizing the properties of the X-ray absorption features.

For the reduction of the \chandra\ observations we used 
standard CXC threads to screen the data for 
status, grade, and time intervals of acceptable aspect solution and background levels.
The pointings placed 
\apm\ on the back-illuminated S3 chip of ACIS.
To improve the spatial resolution we
removed a {$\pm$~0\sarc25} randomization applied to the event positions
in the CXC processing and employed a sub-pixel resolution technique
developed by Tsunemi et al. (2001).

In both the \xmm\ and \chandra\ analyses we tested the sensitivity of our results 
to the selected background and source-extraction
regions by varying the locations of the background regions and varying the 
sizes of the source-extraction regions. We did not find any significant change in the 
background-subtracted spectra.  For all models of  
\apm\ we included Galactic absorption due to neutral gas with 
a column density of  
$N_{\rm H}$=3.9 $\times$ 10$^{20}$~cm$^{-2}$ (Stark et al. 1992). 
All quoted errors are at the 90\% confidence level unless mentioned otherwise.

\subsection{ \chandra\ and \xmm\ Spectral Analysis of  \apm}


We first fitted the \chandra\ and \xmm\ spectra of \apm\ with a simple model consisting 
of a power law with neutral intrinsic
absorption at $z = 3.91$ (model~1 of Table~2). These fits are not acceptable in a statistical sense as 
indicated by the reduced $\chi^{2}$.
The residuals between the fitted simple absorbed power-law model and the data show significant absorption 
for energies in the observed-frame band of $ < $ 0.6~keV (referred to henceforth as low-energy absorption)
and 2--5~keV (referred to henceforth as high-energy absorption).

To illustrate the presence of these low- and high-energy absorption features, we fit the spectra from 
observed-frame 4.5--10 keV with a power-law model (modified by Galactic absorption) and 
extrapolated this model to the energy ranges not fit. The residuals of these fits are shown in Figures 1 and 2.
Significant low- and high-energy absorption is evident in all observations.

We proceed by fitting the spectra of \apm\ with the following models: 
1) absorbed power-law (APL), 
2) absorbed power-law with a notch (APL+No),
3) ionized-absorbed power-law with a notch (IAPL+No),
4) absorbed power-law with an absorption edge (APL+Ed),
5) ionized-absorbed power-law with an absorption edge (IAPL+Ed),
6) absorbed power-law with two absorption lines (APL+2AL),
7) ionized-absorbed power-law with two absorption lines (IAPL+2AL),
8) absorbed power-law with two intrinsic ionized absorbers (APL + 2IA), and
9) absorbed power-law with with two partially covered intrinsic ionized absorbers (APL+PC*(2IA)).
The XSPEC notations for these models are given in the notes of Tables 2 and 3.

The results from fitting these models to the three \xmm\ and two \chandra\ spectra are presented in  
Tables 2 and 3. For spectral fits using models 3, 5, and 7 the low-energy absorption is modeled
using the photo-ionization model \verb+absori+ contained in \verb+XSPEC+ (Done et al. 1992).
We note that the \verb+absori+ model is just
a first approximation to what is likely a more complex situation.
As an independent check of the accuracy of the fits that used the \verb+absori+ model we also
repeated several of these fits using the warm-absorber model \verb+XSTAR+ (Kallman et al. 1996;  Kallman \& Bautista 2001).
\verb+XSTAR+ calculates the physical conditions and emission spectra of photoionized gases.
In the current analysis we use a recent implementation of  the \verb+XSTAR+ model that can be used within
\verb+XSPEC+. 

In most epochs, models APL+2AL  and IAPL+2AL that included two absorption lines provided significant improvements to the fits
compared to models that included an absorption edge. 
Specifically, the $F$-test indicates the significance of the improvements of the IAPL+2AL model 
compared to the IAPL+ Ed model to be  
$\simgt 99.99\%$ ($\Delta\chi^{2} = 42.5$),
$\simgt 84\%$ ($\Delta\chi^{2} = 6.4$),
$\simgt 74\%$ ($\Delta\chi^{2} = 6.3$),
$\simgt 99\%$ ($\Delta\chi^{2} = 15.6$), 
and $\simgt 99.7\%$ ($\Delta\chi^{2} = 17.1$),
for four additional parameters in epochs 1 through 5, respectively.

We note that the X-ray BALs in the 1.5--3.6~keV band (observed-frame) correspond to a significant
detection following the criteria described in \S3 of 
Vaughan \& Uttley  (2008).
Specifically, we find the ratio of the total equivalent
widths\footnote{The equivalent width (EW) is defined as $EW=\int
\frac{F_c-F_E}{F_c}dE$, where $F_c$ is the continuum flux and
$F_E$ is the flux in the absorber.} of the absorption features to
their uncertainty to be $EW/{\sigma_{EW}} \simgt3$ in every
observation (see models 6 and 7 in Table~2).

As indicated in the results of the spectral fits shown in Table 2, the energies and widths
of the absorption lines of models APL+2AL and IAPL+2AL vary significantly between several epochs.
This is also suggested by the residuals shown in Figures 1 and 2, where we indicate 
with arrows the locations of several of the best-fit energies of the absorption lines.

To estimate the significances of variations of the best-fit energies ($E_{\rm abs}$) and widths ($\sigma_{\rm abs}$) 
of the absorption lines between epochs we calculated the $\chi^{2}$ confidence contours of these best-fit parameters.

We first focus on epochs 1 and 5 where the high-energy absorption lines 
appear to be narrower than in other epochs (see Figures 1 and 2)
and also show variability in energy and width.
In Figure 3 we show the 68\%, 90\%, and 99\% $\chi^{2}$ confidence contours of 
$E_{\rm abs}$ versus normalization of the first and second absorption lines 
(APL + 2 AL model) in epoch 1 (top panel) and epoch 5 (bottom panel).
Since the 99\% and 95\% confidence contours of the second absorption line (component $abs2$) between epochs 1 and 5, respectively, do not overlap
in energy we conclude that a significant change occurred at the $ > $ 99.9\% confidence level of the energy of 
the second absorption line between these epochs.


In Figure 4 we show the 68\%, 90\% and 99\% $\chi^{2}$ confidence contours of
 $E_{\rm abs}$ versus $\sigma_{\rm abs}$ of the first and second absorption lines 
(APL + 2 AL model) in epochs 1 and 5. 
The probability that the energy-width parameters of the
first absorption line are the same between epochs 1 and 5 (null probability)
is $\lesssim$$1\times10^{-4}$ 
\footnote{The product of the probabilities of being outside the confidence contours that
barely touch (see Figure~4) is an upper limit to the
null probability}.


An additional significant variation of particular interest is found 
between epochs 3 and 4 that are separated by only 3.3~days
in the rest frame.  The 0.2--10~keV pn count rate of \apm\ varied between epochs 3 and 4 by 36.8 $\pm$ 0.3 \%.
We investigated whether the X-ray BALs varied in this 3.3~day period.
In Figure 5 we show the 68\%, 90\% and 95\% $\chi^{2}$ confidence contours of
$E_{\rm abs}$ versus $\sigma_{\rm abs}$ of the first absorption line (APL + 2 AL model) in epoch 3 (dotted line)  
and epoch 4 (solid line).
The 95\% confidence contours of epoch 3 almost touch 
the 90\% confidence contours of epoch 4.
The probability that the energy-width parameters of the first absorption 
line (model 7; Table~2) are the same between epochs 3 and 4 
(null probability) is $\simlt$ 0.05$\times$0.10 = 5 $\times$ 10$^{-3}$.

In Table 3 we present results from spectral fits
that attempt to model the high-energy absorption with two outflowing ionized absorbers
that are either completely covering the source (model 8) or are partially covering it (model 9).
For spectral fits using models 8 and 9 the high-energy absorption in modeled
using \verb+XSTAR+. 
The low-energy absorption is modeled with a stationary neutral absorber.
For epochs 1 and 5 we find that the combined covering factor
of the highly ionized absorbers is close to unity.
This is expected since the detected ionized absorption lines in epochs 1 and 5 are
narrower than the ones detected in the other epochs whereas the presence of any significant direct component
(emission that does not get absorbed by the outflowing material) would have 
diluted any narrow absorption lines.
For epochs 2, 3 and 4 the spectral fits (model 9 of Table 3) do allow for an emitted component that is not absorbed by the ionized outflowing material.

Our fits with \verb+XSTAR+ indicate an outflowing X-ray absorbing medium with 
ionization parameters ranging between $\log\xi = 2.9$\footnote{Throughout this paper we
adopt the definition of the ionization parameter of Tarter et al. (1969) given by
$\xi=\frac{L_{\rm ion}}{n_H r^2}=\frac{4 \pi}{n_H}
\int_{1Rdy}^{1000Rdy}F_{\nu}d\nu$, where $n_H$ is the hydrogen number
density, and $r$ is the source-cloud separation.} 
and $\log\xi = 3.9$ (model 8 of Table 3).
The two strongest iron lines for this highly ionized absorbing medium have rest (or laboratory)
energies of 6.70 keV (\FeXXV\ $1s^2-1s2p$) and 6.97 keV
(\FeXXVI\ $1s-2p$). In general the \FeXXV\ $1s^2-1s2p$
line will be stronger than the \FeXXVI\ $1s-2p$ line for a
medium with $2.75 \simlt \log\xi \simlt 4.0$ (see Figure 3 of Saez et al. 2009).

We emphasize that the photo-ionization models used in our analysis (i.e., \verb+absori+ and \verb+XSTAR+)
do not consider possible velocity gradients in the outflowing absorber and therefore 
cannot provide realistic models of the X-ray BALs.
We attempt to mimic the velocity broadening of the lines
by introducing in the \verb+XSTAR+ model large turbulent velocities of 
$v_{\rm turb}$ = 10,000~km~s$^{-1}$ for epochs 1 and 5 and $v_{\rm turb}$ = 30,000~km~s$^{-1}$ for epochs 3, 4 and 5.





\section{DISCUSSION}

\subsection{Short- and Long-Term Variability of \apm}
As we showed in \S 2 the 0.2--10~keV pn count-rate of \apm\ varied by $\sim$~36.8 $\pm$ 0.3 \% between epochs 3 and 4. 
From a light-travel time argument we estimate that the observed short-term
variability between epochs 3 and 4 implies a size-scale of the X-ray emission region of 
the order of $l_{\rm var}$ = $c{\Delta}t$/(1 + $z$) $\sim$ 7.4 $\times$ 10$^{15}$~cm.
We compare this emission-region size scale to the radius of the innermost 
stable circular orbit ($r_{\rm ISCO}$) which for Schwarzschild and Kerr (maximally spinning) 
black holes are 6$r_{\rm g}$ and $r_{\rm g}$, respectively, where $r_{\rm g} = GM/c^{2}$ is 
the gravitational radius.
For \apm\ we assume  $M_{\rm BH}\sim 10^{12}M_\odot
\mu_L^{-1}$, where $\mu_L\sim100$ (Egami et al. 2000).\footnote{See,
however, Riechers et al. (2009), who find a magnification of $\mu_L\sim4$.
Riechers et al. (2009) also use the observed width of the CIV line to obtain
a black-hole mass of  \hbox{$M_{\rm BH} \sim 10^{11}
\mu_L^{-1}M_\odot$.}}
We find that $r_{\rm ISCO}$ = 4.5 $\times$ 10$^{15}$~cm for the case of a Schwarzschild black hole
which is comparable to the size-scale of the X-ray emitting region implied by our light-travel time argument.
This result is consistent with recent X-ray and UV (rest-frame) monitoring observations of the $z = 2.32$ gravitationally lensed quasar HE 1104$-$1805 
that have constrained the size of the X-ray emitting region in this object to be smaller than 6$r_{\rm g}$
and the size of the UV emission region to be $\sim$ 30$r_{\rm g}$
based on the analysis of the microlensed light-curves of the lensed images (Chartas et al. 2009).
A hint to the cause of the spectral variability can be seen in Figure 6
where we show the three over-plotted \xmm\ spectra of \apm.  

The spectra between epochs 3 and 4 appear to differ over the entire observed energy band
by approximately a constant. This type of behavior
suggests that the cause of the variability is not the result of a change in the absorbing column density
but is likely due to either a change in covering factor 
or a change in the ionizing flux from the central source.


As shown in Figure 6 the comparison between the epoch 2 and 3 \xmm\ spectra, that are separated by 
$\sim$ 5.4~years (observed-frame), indicates that they converge at energies above $\sim$ 3~keV
but diverge significantly below this value. 
This type of energy-dependent variation is indicative of a decrease in the absorbing column density 
between epochs 2 and 3 and is consistent with the results of our spectral analysis 
shown in Table 2.
Since the relevant energy range ($ > $ 15~keV in the rest frame) is not significantly
affected by neutral or ionized absorption, we interpret the
variation between the epoch 2 and 3 \xmm\ spectra 
as an indication that the intrinsic unabsorbed luminosity of \apm\ has not significantly varied between
these two observations and that most of the observed changes in flux result
from changes in the opacities of the low and/or high-ionization absorbers.

In Figure 7 we over-plot the epoch 1 and epoch 5
{\sl Chandra} spectra that are separated by $\sim$ 5.9~years (observed-frame). 
The spectra appear to converge at energies below $\sim$ 0.7~keV and above $\sim$ 5~keV
and significantly differ between these energies.
As shown in our spectral analysis, the X-ray BALs varied significantly 
between these two epochs which is consistent with the observed difference
of the spectra between 0.7--3.6~keV seen in Figure 7.
The variability of the absorption between epochs 1 and 5 can in general be
caused by at least three factors: (a) transverse motion of the absorber across the
continuum source, (b) changes in the opacity of the absorber, and (c) changes in the velocity
structure of the absorber. Epochs 1 and 5 are separated by 2150~days
which is significantly longer than the expected timescale of
variability due to transverse motion and also longer than the expected flow timescale.
We therefore cannot reach a firm conclusion regarding the cause of the
variability between epochs 1 and 5.  The significant differences in timescales,
however, imply that it is unlikely that the variability is being
produced by the same absorber moving across our line of sight.


To better illustrate the flux variability of \apm\ we show 
the ratios of the observed and best-fit modeled (using model 6 of Table 2) flux densities $F_{\rm i}/F_{\rm j}$
between the three \xmm\ observations in Figure 8 and 
between the two \chandra\ observations in Figure 9, where $i$ and $j$ represent the 
epochs compared, $F$ is the flux density in units of counts~s$^{-1}$~keV$^{-1}$ 
and ${\Delta}t$ is the observed-frame time between epochs.
We first focus on the short-term variability between epochs 3 and 4 shown in Figure 8.
We notice that the variability of the flux density is to first order independent of energy 
over the observed 0.5--10~keV band with the exception of 
apparent variability in the region of 1.5--3.6~keV (observed-frame) near the Fe BALs.
A plausible explanation of the energy independence of the short-term variability of the flux density
may involve motion of the highly ionized absorbing outflow
perpendicular to our line of sight.This motion may change the covering factor resulting in  
possible energy-independent variability of the flux density. A second possible explanation is a change in the ionizing flux 
of the central source. Our spectra fits that assumed a model that includes emission
through two ionized absorbers and emission that
is not absorbed by the ionized absorber (model 9 of Table 3)
suggest a possible decrease in the total covering factor of the ionized absorbers between 
epochs 3 and 4. Unfortunately the limited S/N of the current spectra provide
poor constraints on the covering factors and 
therefore the implied change in covering factor between epochs 3 and 4 can only be considered suggestive.

We second center on the long-term variability between epochs 2 and 3 shown in Figure 8.
The variability of the flux density between these epochs becomes
large below $\sim$ 2~keV, however, no significant variability is 
detected above this energy.  A possible explanation of such energy dependent 
variability is a decrease in column density between epochs 2 and 3.

We finally comment on the long-term variability between epochs 1 and 5 shown in Figure 9.
We note that the flux density appears to have varied by a larger factor near
the region of the Fe BALs  that extends above $\sim$ 1.5 keV (observed frame).
A possible explanation for this flux-density variability between these epochs is a change in the 
properties of the outflowing ionized absorbers. 
This interpretation is consistent with the results of our previous spectral analysis that showed a significant 
change in the energies and widths of the absorptions lines (see Figures 3 and 4).

\subsection{Mass-Outflow Rate and Efficiency of the Outflow}

Our {\it XMM-Newton} and {\it Chandra} observations of \apm\ provide constraints 
on the velocity, column density, and location of the X-ray BAL
material that allow us to estimate the mass-outflow rate from the expression:

\begin{equation}
 \dot{M_{\rm i}} = 4{\pi}r_{\rm i}(r_{\rm i}/{\Delta}{r_{\rm i}})N_{\rm H,i}m_{p}v_{wind,i}f_{c,i}. 
\end{equation}

\noindent
where ${\Delta}{r_{\rm i}}$ is the thickness of the absorber at radius $r_{\rm i}$, 
$N_{\rm H,i}$ is the hydrogen column density,
$v_{\rm wind,i}$ is the outflow velocity of the X-ray absorber, $f_{\rm c,i}$
is the global covering factor of the absorber, 
and $i$ indicates the absorbing component. 
Combining the constraints on the mass-outflow rate and the bolometric luminosity
we can estimate the efficiency of the quasar outflow. The efficiency is defined as 
the ratio of the rate of kinetic energy injected into the interstellar medium (ISM) and intergalactic medium (IGM)
by the outflow to the quasar's bolometric luminosity, i.e.,
 
\begin{equation}
\epsilon_{\rm k,i} = {{1}\over{2}}{\dot{M}_{\rm i}{v^{2}_{\rm wind,i}}\over{L_{\rm Bol}}} = 2 {\pi}{f_{\rm c,i}}r_{\rm i}(r_{\rm i}/{\Delta}{r_{\rm i}})N_{\rm H,i}m_{\rm p}{{v^{3}_{\rm wind,i}}\over{L_{\rm Bol}}}
\end{equation}

\noindent
where $L_{\rm Bol}$ is the bolometric luminosity of the quasar.


The best-fit values of the equivalent widths of the X-ray absorption lines (assuming model 6 of Table 2) were
used to estimate the hydrogen column densities $N_{\rm H}$ of the X-ray BALs 
using a curve-of-growth analysis (e.g., Spitzer 1978).
For our crude calculation we assumed 
$b$ parameters of the order of the observed widths of the lines
where, $b = \sqrt{2}{\sigma_{u}}$ and $\sigma_{u}$ is the velocity width of the line.
The bulk outflow velocities of each outflow component were inferred from the best-fit values of the 
energies of the blueshifted absorption lines obtained by spectral fits to 
the data assuming model 6 of Table 2.
To convert from line energies to outflow velocities we used the relativistic Doppler formula
assuming that the high-energy absorption lines are the result of resonance 
absorption from ions of \FeXXV\ in a gas with solar abundances, and the angle between our line of sight
and the outflow direction is zero.
For estimating the outflow efficiency of \apm\ we assumed $L_{\rm bol}=7\times10^{15}\mu_L^{-1}L_\odot$ 
(Irwin et al. 1998; Reichers et al. 2009). We assume a conservatively wide range for the covering factor of $f_{\rm c}$=0.1--0.3 
based on the observed fraction of BAL quasars (e.g., Hewett \& Foltz 2003).
We obtained a constraint on the covering fraction of the flow by placing limits
on the presence of any emission components of the lines.
Crudely the ratio of the emission to absorption equivalent widths is proportional
to the global covering factor $f_{\rm c}$. The tightest upper bound is
found for the \xmm\ observation of \apm\ in epoch 3. We find 
that $EW_{\rm em1}$/$EW_{\rm abs1}$ $ < $  0.24 at the 
90\% confidence level assuming model 6 of Table 2 and we expect
$f_{\rm c}$ $\propto$ $EW_{\rm em1}$/$EW_{\rm abs1}$.
We note that the covering factor $f_{\rm c}$ will depend on the fluorescence yields of the ionized absorber which
for \FeXXV\ and \FeXXVI\  are 0.5 and 0.7, respectively ( e.g., Krolik \& Kallman 1987). 
For this crude estimate we included a narrow emission line in model 6
at a rest-frame energy of 6.7 keV. The exact 
relevant energy of the emission line to consider is probably geometry dependent. 
For estimating the outflow efficiency we assumed a fraction $r/{\Delta}{r}$ ranging from 1--10 based on theoretical models of quasar outflows (e.g., Proga \& Kallman 2000). 
Assuming that the maximum outflow velocity is
produced by gas that has reached its terminal velocity
one obtains the approximation $R_{\rm launch} \sim R_{\rm s}(c/v_{\rm wind})^{2}$, 
where $v_{\rm wind}$ is the observed outflow velocity.
Based on our estimated maximum outflow velocity ($v = 0.76~c$) we expect $r$ to be similar
to $R_{\rm launch}$ and range between 3~$R_{\rm s}$  and 15~$R_{\rm s}$ (see Figure 4 of Saez et al. 2009).
We note that the short term variability time-scale of the X-ray BALs (3.3~d in rest-frame) is consistent with a launching radius 
of a few times $R_{\rm S}$.

We used a Monte Carlo approach to estimate the errors ofÊ
$\dot{M}_{\rm i}$ and $\epsilon_{\rm k}$.
The values of $f_{\rm c}$, $r/{\Delta}{r}$, and $r_{\rm i}$ were assumed to have uniform distributions within their error limits.
The values of $N_{\rm H}$ were assumed to have normal distributions within their error limits.
By multiplying these distributions and with the appropriate constants from equations 1 and 2 we obtainedÊ
the distributions of $\dot{M}_{\rm i}$ and $\epsilon_{\rm k}$.
We finally determined the means of the distributions of 
$\dot{M}_{\rm i}$ and $\epsilon_{\rm k}$ and estimated the 68\% confidence ranges.

In Table~4 we present the outflow velocities, the column densities, the mass-outflow rates
and the efficiencies of the outflowing components.

\subsection{Driving Mechanism of X-ray BALs}
One of the major unanswered questions in current theoretical and numerical models 
of quasar winds involves explaining how highly ionized outflows of X-ray absorbing 
material become accelerated to near-relativistic velocities. The main problem is that for 
radiatively driven winds the magnitude of the force on an absorber is a function of its 
ionization state. As the ionization parameter of the absorber increases, 
fewer resonant transitions are available to absorb photons from the source, thus resulting in a weaker driving force. Chelouche \& Netzer (2001) have explored the dependence of the
radiation driving force on the ionization parameter of the outflowing gas
and they conclude that for a highly ionized X-ray absorbing gas, the average force
multipliers are close to 10.
  
Most current theoretical work on radiative acceleration in quasars 
has focused on interpreting the outflows of UV absorbers 
(e.g, Murray et al. 1995; Proga et al. 2000; Proga et al. 2004; Everett 2005), 
however; there are no self-consistent models that can 
explain the acceleration mechanism that 
leads to the near-relativistic outflows of X-ray absorbing material.

A clue to understanding the acceleration process of the wind in \apm\
is perhaps provided by the observed large difference
between the maximum velocities of the UV ($v_{\rm UV}$ $\sim$ 0.04$c$; e.g., Srianand \& Petitjean 2000) 
and X-ray ($v_{\rm X-ray}$ $\sim$ 0.43--0.76$c$) absorbers.
One possible explanation of the difference between the UV and X-ray outflow velocities 
is that the UV and X-ray BALs are produced by 
different absorbers, with UV emission from the accretion disk driving the UV absorbers
and X-ray emission from the hot corona contributing to the acceleration of the X-ray absorbers.
If the X-ray absorbers are partly driven by radiation from the hot corona we might expect to 
detect a correlation between the properties of the X-ray BALs 
and the properties of the X-ray spectrum. 
For example, we might expect the maximum outflow velocity of the X-ray absorbers
to depend on the shape of the X-ray spectrum that at first order can be represented with the X-ray photon index ($\Gamma$).

The minimum and maximum projected velocities ($v_{\rm
min}$, $v_{\rm max}$) of the outflow are estimated from the
minimum and maximum energy ranges ($E_{\rm min},E_{\rm max}$) of
the high-energy absorption features in \apm\ assuming that the high-energy X-ray BALs are produced by the 
resonance transition \FeXXV\ ($1s^2-1s2p$). We obtained $E_{\rm
min}$ and $E_{\rm max}$ from our spectral fits assuming the
two absorption-line (APL+2AL) model.
Specifically, based on the best-fit values of the
APL+2AL model (model 6; Table 2), we obtain $E_{\rm
min}=E_{\rm abs1}-2\sigma_{\rm abs1}$ and $E_{\rm max}=E_{\rm
abs2}+2\sigma_{\rm abs2}$. 
The values of $E_{\rm min}$ and $E_{\rm max}$ are
presented in Table~5.

In Figure 10 we show the maximum outflow velocity observed from \apm\ as a function of X-ray 
photon index based on the {\sl XMM-Newton}, {\sl Chandra}, and {\sl Suzaku} observations. 
We notice a possible trend between the photon index and 
the maximum outflow velocity in the sense that flatter X-ray spectra
appear to result in lower outflow velocities.
One possible explanation is that flatter X-ray spectra over-ionize the 
X-ray absorber resulting in a decrease of the force multiplier and a lower outflow velocity.
If this trend is confirmed with future observations it would imply that
the dominant acceleration mechanism responsible for the near-relativistic velocities 
of the X-ray outflowing gas in BAL quasars is indeed radiation driving
as opposed to magnetic driving.

We investigated our hypothesis of the origin of the possible ${\Gamma}-v_{\rm max}$ correlation
by calculating the force multiplier as a function of the incident 
spectral energy distribution (SED). 
The force multiplier represents the ratio by which the 
bound-bound (line) and bound-free (continuum) opacity increases 
the radiation force relative to that produced by Thomson scattering alone. 

We performed calculations of force multipliers
assuming a thin slab illuminated by an ionizing continuum using the CLOUDY
code. We compared our results to those of Arav et al. (1994) and Everett et al. (2005)
who have used a similar approach to ours. In general our computations are in good agreement.
We performed calculations assuming the SED is
a power-law extending from the UV (or 1~Ryd $\sim$13.6 eV)
to hard X-rays (or $10^4$~Ryd $\sim$100 keV). For our calculations
we assume SEDs with two different values of the power-law photon
index. A ``soft'' ($\Gamma=2.1$) and a ``hard'' ($\Gamma=1.7$)
SED.
For each SED we calculated the continuum
($M_{\rm C}$) and the line ($M_{\rm L}$) components of the force multiplier.
$M_L$ depends on an additional parameter, $t$, \footnote
{The dimensionless optical depth is $t = n_e \sigma_T v_{\rm th}/(dv/dr)$,
where, $n_{\rm e}$ is the electron number density,  $\sigma_{\rm T}$ is the
Thomson cross section and  $v_{\rm th}$ is the thermal velocity of the gas.
The line force multiplier increases with decreasing $t$.}
which is commonly referred to as the ``effective electron optical depth'' and
encodes the dynamical information of the wind in the radiative
acceleration calculation (see equation 2.5 of Arav \& Li 1994). 
For our calculations we have assumed ${\rm log}~t=-7$.
In Figure 11 we show the $M_{\rm C}$ and $M_{\rm L}$ components of the force multiplier
as a function of the ionization parameter and 
for the soft and hard SED cases.
Our simple ionized absorber model indicates that an increase
of the photon-index from 1.7 to 2.1 of the incident spectrum 
will result in a large increase of the force multiplier.
This result is consistent with the possible trend between $\Gamma$ and $v_{\rm max}$
shown in Figure 10.  

Theoretical models of AGN outflows (e.g., Murray et al. 1995; Chelouche \& Netzer 2003) and
our observations of \apm\ suggest that the 
UV BAL gas could be shielded from the driving radiation 
by an absorbing medium surrounding the central source.
We included the effect of such a shield on our calculated values of the 
force multiplier by assuming that the soft and hard SEDs are attenuated by an
absorber with $\lnh =22.8$ and the temperature at the illuminated
face of the absorbing shield has the value log~$T[K]$ = 5.0
(log~$\xi$~$\sim$~1.4).  As shown in Figure 11, $M_{\rm C}$ decreases in the absorbed-SED case for log~$\xi$~$\simlt$~2,
whereas, $M_{\rm L}$ increases in the absorbed-SED case.
We find that $M_{\rm L}$ increases by almost an order of magnitude
in the absorbed-SED case when the photon index
increases from 1.7 to 2.1.
In Figure 11 we show that in addition to the SED and t parameter, the force
multiplier depends strongly on the ionization parameter.
In Figure 11b the force multiplier for a shielded outflow
drops from about 1000 to 500 between log~$\xi$ of 2--2.8 and
drops from about 500 to 1 between log~$\xi$ of 2.8--3, i.e., similar to the
observed range of the ionization parameter of \apm.
We expect that we observe the ionized X-ray absorber of \apm\ while it has
obtained its terminal velocity.
The observed large value of the ionization parameter of
\apm\ may therefore not be representative of the
ionization parameter during the initial acceleration phase of the absorber.

We note that our assumption that the driving force on the high-energy absorbers is produced primarily
by X-rays is plausible since the short term variability time-scale of the X-ray BALs of \apm\ suggest
a launching radius of $\lesssim$ $10 R_{\rm S}$ (e.g, Chartas et al. 2002; Saez et al. 2009; this work)
and recent studies of AGN employing the microlensing technique indicate that the
X-ray emission region of the hot corona in AGNs is 
compact with a half-light radius of a few $R_{\rm S}$ and their UV regions 
are roughly a factor of ten larger (e.g, Morgan et al 2008; Chartas et al. 2009).
Therefore UV radiation is not expected to contribute initially at small radii
to driving the X-ray absorbing outflow.

If the $\Gamma-v_{\rm max}$ trend is confirmed with additional observations we plan to 
produce a more sophisticated model that will include
more realistic kinematic, ionization and absorption properties of the outflow.


\subsection{Constraints on Outflow Direction}
The observed maximum projected outflow velocities of the X-ray BALs
combined with the constraint that the
maximum velocity of the X-ray absorber along the outflow direction must satisfy  $v_{\rm outflow} < c$
leads to a constraint on the maximum angle between the outflow direction and 
the observed line of sight through the absorber($\theta_{\rm max}$).
The best-fitted values of the ionization parameter for the outflowing absorbers (see model 8 of Table~3) 
lie in the range of $2.9 \simlt \log\xi \simlt 3.9$ and imply that the high-energy absorption is the result of resonance 
absorption from ions of \FeXXV\ and/or \FeXXVI.
Since the present X-ray spectra cannot distinguish between these ionization stages
we estimate the angle between the outflow direction and our line of sight 
considering that the X-ray BALs are produced by resonance transitions of either \FeXXV\ ($1s^2-1s2p$),  
\FeXXV\ ($1s^2-1s3p$), \FeXXVI\ ($1s-2p$) and \FeXXVI\ ($1s-3p$).
In Figure 12 we show the outflow velocity versus the angle between
the direction of the outflowing absorber 
and our line of sight through the absorber.
The outflow velocities were calculated using the relativistic Doppler formula 
for the case of the maximum rest-frame energy of the X-ray BALs of $E_{max}$ = 17.9~keV (see Table ~5)
observed in epoch 3.
For the case where the X-ray BALs are produced by the resonance transition of \FeXXV\ ($1s^2-1s2p$)
the detected projected maximum outflow velocity of $v_{\rm max} \sim 0.76c$ 
leads to the constraint of $\theta_{\rm max}$  $\simlt$ 22$^{\circ}$.
Such a small angle is consistent with the unification scheme of BAL and non-BAL quasars
that posits that BAL quasars are viewed 
almost along the outflow direction.

\section{CONCLUSIONS}

We have presented results from an analysis of three \xmm\ and two \chandra\ observations of 
the $z = 3.91$ BAL quasar \apm.
The main goals of these observations were to study the kinematic and photoionization 
properties of the wind in order to assess whether it plays an important role in 
controlling the evolution of the host galaxy and central black hole. 
Additional objectives included understanding the mechanism 
that drives the X-ray absorbers to near-relativistic velocities and constraining the 
geometry of the outflow.

The main conclusions of our spectral and timing analyses are the following:

(a) X-ray BALs lying in the range of 1.5--3.6~keV (observed frame) 
are detected at the greater than 99.9\% confidence level
in all five observations of \apm. We note that a recent analysis of three {\sl Suzaku} observations
of \apm\ also shows significant detections of X-ray BALs in all three observations
(Saez et al. 2009).

(b) Based on our spectral analysis we identify the following components of the outflow.
First, a highly ionized X-ray absorbing material with an ionization parameter in the range 
of $2.9 \simlt \log\xi \simlt 3.9$ and a column density of  $\log{N_{\rm H}} \sim 23$ 
outflowing at velocities of up to 0.76~$c$, and second a low-ionization X-ray absorbing gas with a 
column density of $\log{N_{\rm H}} \sim 22.8$  that may constitutes the
X-ray shielding gas that is thought to protect the UV absorber from becoming over-ionized.
Models that include partial covering result in even larger column densities of the ionized absorber of up to $\log{N_{\rm H}} \sim 24$, however, 
we caution that these values are not well constrained (see model 9 of Table~3) by the spectral fits.
We note that no iron overabundance is required to fit the X-ray spectra 
of \apm\ consistent with the spectral analysis of previous observations of this object.

(c) Significant variability of the X-ray continuum and X-ray BALs over short and long time-scales is detected. 
Specifically, we detect a  $\sim$ 36.8 $\pm$ 0.3 \% change of the 0.2--10~keV pn count-rate 
between epochs 3 and 4 that are separated by only 
3.3~days (proper-time). Based on a simple light-travel time argument
this variability time-scale implies a radius of the emission region
of $\sim$ 7.4 $\times$ 10$^{15}$~cm which is comparable 
to $r_{\rm ISCO}$ = 4.5 $\times$ 10$^{15}$~cm for the case of a Schwarzschild black hole in \apm.
We speculate based on the energy independent spectral change 
between epochs 3 and 4 (with the exception of the Fe BAL region) 
that the cause of this variability is a change in the covering fraction as the
outflowing absorbing material moves across our line of sight.
Significant changes in the flux densities between epochs 2 and 3 
that are separated by $\sim$ 5.4 years (observed-frame) are detected.
The variation of flux density is significant below 3~keV (observed frame)
showing a increase toward lower energies but is near zero above 3~keV.
This type of energy-dependent change 
suggests that the intrinsic unabsorbed luminosity of \apm\ has not significantly varied between these epochs
and the long-term variability is the result 
of a decrease in the opacity of the absorber. This interpretation is consistent with the results of our spectral analysis.

(d) Assuming our interpretation that the high-energy X-ray BALs are produced by resonance transitions of 
highly-ionized iron 
we estimate the mass-outflow rate 
and efficiency of the outflow to have varied over a period of 1.2 years (proper-time)
between $16_{-8}^{+12}$ $M_{\odot}$~yr$^{-1}$ and 
$64_{-40}^{+66}$ $M_{\odot}$~yr$^{-1}$ and $0.18_{-0.11}^{+0.15}$  to $1.7_{-1.2}^{+1.9}$, respectively.

(e) The detected projected maximum outflow velocity of $v_{\rm max} \sim 0.76c$ 
leads to a constraint on the angle between
the outflow direction and 
the observed line of sight through the absorber of
$\theta_{\rm max}$ $\simlt$ 22$^{\circ}$ (assuming \FeXXV($1s^2-1s2p$))
and $\theta_{\rm max}$ $\simlt$ 26$^{\circ}$ (assuming \FeXXVI($1s-2p$)).
Such a small angle is consistent with the unification scheme of BAL and non-BAL quasars
that posits that BAL quasars are viewed 
almost along the outflow direction.

(f) A possible trend is found between the X-ray photon index and 
the maximum outflow velocity of the ionized X-ray absorber
in the sense that flatter X-ray spectra appear to result in lower outflow velocities.
One possible explanation is that flatter X-ray spectra over-ionize the 
X-ray absorber resulting in a decrease of the force multiplier of the X-ray absorber 
and thus a lower outflow velocity.

\acknowledgments
We acknowledge financial support from NASA via the Smithsonian Institution grant SAO SV4-74018
and from NNX08AB71G. WNB acknowledges financial support from NASA LTSA grant NAG5-13035.
GC thanks the inspirational environment offered by Saint's Cafe
and also thanks the members of the Penn State outflow journal club for many insightful discussions
on quasar winds.

\clearpage

\clearpage
\begin{table}
\caption{Log of Observations of \apm}
\small
\begin{center}
\begin{tabular}{lccccccc}
 & & & & &&&\\ \hline\hline
                  &                      &                     &   &    &  & &\\
Observation Date  & Observatory    &  Observation  & Time$^{a}$  & $N_{\rm sc}$$^{b}$  & $f_{0.2-2}$$^{c}$& $f_{2-10}$$^{c}$ \\
           &                      &  ID                  & (ks)     &  net counts & erg~s$^{-1}$~cm$^{-2}$ & erg~s$^{-1}$~cm$^{-2}$\\
\hline
\hline

2002 February 24 (Epoch 1)          & {\it Chandra}            & 2979             &    88.82  & 5,627  $\pm$ 75       &  1.8$_{-0.1}^{+0.1}$ &  4.3$_{-0.1}^{+0.1}$  \\
2002 April 28 (Epoch 2)        & {\it XMM-Newton}   & 0092800201   &  83.46    & 12,820  $\pm$ 139    &  1.9$_{-0.1}^{+0.1}$ &  4.1$_{-0.1}^{+0.1}$\\
2007 October 06  (Epoch 3)          & {\it XMM-Newton}    & 0502220201	&  56.38   & 11,400 $\pm$ 114     &  2.5$_{-0.1}^{+0.1}$ &  3.9$_{-0.1}^{+0.1}$\\
2007 October 22 (Epoch 4)           & {\it XMM-Newton}    & 0502220301   & 60.37    &  16,698 $\pm$ 133    &  3.5$_{-0.1}^{+0.1}$ &  5.0$_{-0.1}^{+0.1}$ \\
2008 January 14 (Epoch 5)           &{\it Chandra}               &  7684            &  88.06   & 6,938  $\pm$  83      &  1.9$_{-0.2}^{+0.2}$  &  4.5$_{-0.2}^{+0.2}$ \\



\hline \hline
\end{tabular}
\end{center}
\tablenotetext{a}{Time is the effective exposure time remaining after the application of good time-interval (GTI)
tables and the removal of portions of the observation that were severely contaminated by background flaring.}
\tablenotetext{b}{Background-subtracted source counts including events with energies within the 0.2--10~keV band.
The source counts and effective exposure times for the \xmm\ observations refer to those obtained with the
EPIC PN instrument.
See \S 2 for details on source and background extraction regions used for measuring $N_{\rm sc}$.}
\tablenotetext{c} {The absorbed fluxes (in units of 10$^{-13}$~ergs~cm$^{-2}$~s$^{-1}$) in the 0.2--2~keV and 2--10~keV observed-frame band are obtained using the 
model APL+2AL (model 6; \S 3). The errors are at the 68\% confidence level.}

\end{table}

\clearpage
\begin{deluxetable}
{ccccccccc} \tabletypesize{\small}
\tablecolumns{18} \tablewidth{0pt} \tablecaption{ Results from
spectral fits to the \chandra\ and \xmm\ observations of \apm.
\label{tab:modn}}

\tablehead{
\colhead{Model$^a$} & \colhead{Parameter$^c$} & \colhead{Values Epoch 1$^d$} & \colhead{Values Epoch 2$^d$} & \colhead{Values Epoch 3$^d$} &  \colhead{Values Epoch 4$^d$} & \colhead{Values Epoch 5$^d$}  }

\startdata

1....... & $\Gamma$           & $1.74_{-0.06}^{+0.06}$  & $1.94_{-0.05}^{+0.05}$  & $2.11_{-0.05}^{+0.05}$ & $2.14_{-0.03}^{+0.04}$ & $1.94_{-0.05}^{+0.07}$ \\
&  $N_{\rm H}$     & $4.23_{-0.73}^{+0.78}$  & $5.9_{-0.49}^{+0.52}$   & $5.26_{-0.38}^{+0.40}$  & $5.50_{-0.30}^{+0.31}$ & $10.5_{-1.58}^{+1.27}$ \\
&   $\chi^2/\nu$      &  189.8/109                                              & 192.8/120                       & 191.1/110                         & 244.8/148  & 100.9/75     \\
&   $P(\chi^2/\nu)$$^e$ &  $2.6\times10^{-6}$        & 0.025                             & $2.8\times10^{-5}$         &$2.6\times10^{-6}$           &  $9.7\times10^{-7}$   \\
\\
2....... &  $\Gamma$            & $1.75_{-0.02}^{+0.02}$  & $1.87_{-0.05}^{+0.04}$        & $2.04_{-0.04}^{+0.04}$    & $2.08_{-0.03}^{+0.04}$ & $1.95_{-0.04}^{+0.04}$ \\
&  $N_{\rm H}$                   &  $4.6_{-0.77}^{+0.84}$     & $6.19_{-0.45}^{+0.51}$      & $5.39_{-0.38}^{+0.39}$     & $5.64_{-0.31}^{+0.29}$ & $10.85_{-0.96}^{+1.30}$  \\
&  $E_{\rm notch}$[keV]      & $8.04_{-0.07}^{+0.10}$        & $11.24_{-0.72}^{+0.50}$      & $11.62_{-0.39}^{+0.44}$ & $12.91_{-0.91}^{+1.30}$ & $8.49_{-0.2}^{+0.2}$ \\
&  ${\rm W_{notch}}$[keV] & $0.23_{-0.06}^{+0.06}$   & $7.70_{-1.50}^{+0.90}$       & $8.87_{-0.79}^{+1.28}$   & $11.85_{-1.90}^{+2.29}$ & $0.16_{-0.06}^{+0.07}$ \\
&  ${\rm f_{notch}}$[keV]  & $1$                                                                 & $0.21_{-0.04}^{+0.04}$       & $0.21_{-0.04}^{+0.04}$   &  $0.16_{-0.03}^{+0.03}$  &  $1$ \\
&   $\chi^2/\nu$                    &  151/107                                                   & 122.9/117                                       & 117.4/107                         & 175.8/145  & 83.2/73 \\
&   $P(\chi^2/\nu)$               & $3.2\times10^{-3}$                                      & 0.34                                           & 0.23                                   & 0.04 & 0.19     \\

\\
3....... &  $\Gamma$              & $1.78_{-0.03}^{+0.06}$   & $1.92_{-0.05}^{+0.05}$        & $2.08_{-0.03}^{+0.05}$    & $2.10_{-0.04}^{+0.04}$  & $1.94_{-0.05}^{+0.05}$\\
&  $N_{\rm H}$                   &  $15.6_{-8.6}^{+8.3}$     & $13.5_{-4.49}^{+4.85}$      & $9.92_{-4.1}^{+4.0}$           & $8.36_{-3.45}^{+2.18}$  & $11.61_{-1.58}^{+1.13}$ \\
&   $\xi$                              &  $<$0.18                                                &  $150_{-127}^{+120}$               &  $86_{-86}^{+192}$          &  $32_{-12}^{+97}$  & $ < $ 9 \\
&  $E_{\rm notch}$[keV]      & $8.0_{-0.1}^{+0.1}$            & $11.15_{-0.77}^{+0.69}$      & $11.54_{-0.36}^{+0.60}$    & $12.82_{-0.53}^{+0.77}$  & $8.54_{-0.2}^{+0.2}$\\
&  ${\rm W_{notch}}$[keV] & $0.25_{-0.06}^{+0.06}$    & $8.12_{-1.59}^{+1.42}$       & $8.94_{-0.73}^{+1.17}$       & $11.78_{-1.75}^{+2.29}$ & $0.14_{-0.07}^{+0.07}$\\
&  ${\rm f_{notch}}$[keV]  & $1$                                                                  & $0.18_{-0.04}^{+0.04}$       & $0.20_{-0.04}^{+0.04}$       &  $0.16_{-0.03}^{+0.03}$  &  $1$\\
&   $\chi^2/\nu$                    &  146/106                                                    & 112.7/116                            & 114.1/106                              & 175.4/144  & 79.5/72 \\
&   $P(\chi^2/\nu)$               & $6.1\times10^{-3}$                                         & 0.57                                   & 0.28                                       & 0.04  & 0.25  \\
\\

4.......   &$\Gamma$              & $1.72_{-0.06}^{+0.06}$     & $1.88_{-0.05}^{+0.05}$    & $2.07_{-0.04}^{+0.04}$ & $2.12_{-0.03}^{+0.03}$ & $1.95_{-0.06}^{+0.06}$ \\
&  $N_{\rm H}$                     & $4.78_{-0.70}^{+0.86}$       &  $6.22_{-0.46}^{+0.53}$ & $5.43_{-0.39}^{+0.40}$ & $5.74_{-0.31}^{+0.32}$ &$11.26_{-1.11}^{+1.17}$ \\
&  $E_{\rm zedge}$[keV]       & $7.66_{-0.18}^{+0.17}$     & $7.57_{-0.16}^{+0.16}$    & $7.33_{-0.17}^{+0.18}$ & $7.10_{-0.13}^{+0.15}$ & $7.60_{-0.28}^{+0.25}$ \\
&  ${\rm \tau_{zedge}}$        & $0.40_{-0.11}^{+0.11}$      & $0.43_{-0.`09}^{+0.09}$    & $0.38_{-0.09}^{+0.09}$ & $0.30_{-0.07}^{+0.07}$  & $0.21_{-0.09}^{+0.09}$\\
&   $\chi^2/\nu$                      &  144.7/107                                                    & 121.9/118                           & 133.9/108                             & 186.4/146      & 83.4/73                 \\
&   $P(\chi^2/\nu)$                &  $8.9\times10^{-3}$                                         & 0.38                                   & 0.05                                       & 0.01 & 0.19  \\

\\
5.......   &$\Gamma$               & $1.72_{-0.05}^{+0.06}$     & $1.94_{-0.05}^{+0.05}$    & $2.13_{-0.05}^{+0.04}$       & $2.16_{-0.04}^{+0.04}$ & $1.95_{-0.07}^{+0.08}$ \\
&  $N_{\rm H}$                    & $4.6_{-0.7}^{+4.1}$             &  $15.69_{-5.4}^{+5.7}$ & $12.3_{-4.11}^{+3.94}$           & $10.37_{-2.66}^{+4.02}$ &$13.5_{-3.66}^{+5.33}$ \\
&   $\xi $                                 &  $<$77                                                        & $242_{-182}^{+315}$        & $194_{-159}^{+252}$          &  $93_{-76}^{+209}$  & $ < 28$ \\
&  $E_{\rm zedge}$[keV]       & $7.68_{-0.11}^{+0.16}$     & $7.30_{-0.23}^{+0.29}$    & $7.17_{-0.09}^{+0.16}$         & $7.02_{-0.12}^{+0.11}$ & $7.08_{-0.78}^{+0.89}$ \\
&  ${\rm \tau_{zedge}}$        & $0.38_{-0.10}^{+0.11}$      & $0.35_{-0.08}^{+0.07}$    & $0.36_{-0.08}^{+0.09}$         & $0.31_{-0.07}^{+0.07}$  & $0.14_{-0.09}^{+0.09}$\\
&   $\chi^2/\nu$                     &  144.5/106                                                     & 112.8/117                         &  126.4/107                             & 178.9/145       & 83.0/72                  \\
&   $P(\chi^2/\nu)$                &  $7.8\times10^{-3}$                                          & 0.59                                   & 0.10                                      & 0.03  & 0.18  \\
\\

6.......  &$\Gamma$                 & $1.74_{-0.05}^{+0.06}$       & $1.89_{-0.05}^{+0.05}$     &  $2.03_{-0.06}^{+0.05}$  & $2.11_{-0.03}^{+0.04}$  & $1.94_{-0.06}^{+0.06}$ \\
&   $N_{\rm H}$                      &  $4.89_{-0.73}^{+0.80}$     &  $6.29_{-0.52}^{+0.55}$  & $5.41_{-0.39}^{+0.41}$    & $5.74_{-0.31}^{+0.32}$   & $11.35_{-0.76}^{+1.18}$\\
&  $E_{\rm abs1}$[keV]           &  $8.05_{-0.07}^{+0.09}$     & $8.16_{-0.23}^{+0.30}$     & $8.28_{-0.38}^{+0.41}$    & $7.56_{-0.16}^{+0.17}$  & $8.32_{-0.24}^{+0.22}$ \\
&  $\sigma_{\rm abs1}$[keV]    & $ < $ 0.15                          &  $0.31_{-0.31}^{+0.38}$   &  $0.93_{-0.32}^{+0.40}$    &   $0.31_{-0.14}^{+0.17}$  & $0.48_{-0.17}^{+0.27}$     \\
&  EW$_{\rm abs1}$[eV]$^b$ &  $248_{-104}^{+103}$                 & $380_{-298}^{+248}$        & 488$_{-451}^{+447}$        & $189_{-106}^{+110}$  &  $333_{-25}^{+24}$  \\
&  $E_{\rm abs2}$[keV]           &  $9.78_{-0.18}^{+0.19}$   & $10.94_{-0.73}^{+1.15}$    & $12.04_{-3.10}^{+1.29}$   & $10.41_{-0.70}^{+0.80}$  & $13.51_{-0.73}^{+1.59}$\\
&  $\sigma_{\rm abs2}$[keV]    & $0.43_{-0.15}^{+0.17}$     & $2.05_{-0.53}^{+0.79}$     & $2.95_{-1.51}^{+3.46}$      &   $2.95_{-0.74}^{+1.29}$    & $0.92_{-0.92}^{+2.29}$        \\
&  EW$_{\rm abs2}$[eV]$^b$  &  $461_{-103}^{+104}$              & $1008_{-707}^{+446}$        & 1643$_{-1146}^{+506}$   & 864$_{-340}^{+313}$   & $499_{-320}^{+322}$         \\
&   $\chi^2/\nu$                        &  106.1/103                                                & 121.8/114                          & 121.0/104                              & 179.0/142     & 64.9/69              \\
&   $P(\chi^2/\nu)$                   & 0.40                                                                & 0.29                                 & 0.17                                     & 0.02        & 0.62                      \\

\\
\\
7.......  &$\Gamma$                   & $1.74_{-0.06}^{+0.05}$    & $1.96_{-0.06}^{+0.05}$     &  $2.03_{-0.06}^{+0.07}$    & $2.16_{-0.04}^{+0.05}$  & $1.92_{-0.06}^{+0.06}$\\
&   $N_{\rm H}$                      &  $4.7_{-0.7}^{+1.9}$             &  $20.7_{-6.3}^{+6.9}$       & $5.80_{-0.08}^{+0.08}$      & $13.32_{-3.29}^{+3.83}$  & $11.5_{-1.0}^{+5.4}$ \\
&   $\xi $                                 &  $<$77                                                   &  $498_{-321}^{+474}$           & $<$0.22                               &  $228_{-95}^{+84}$ & $ < 36$ \\
&  $E_{\rm abs1}$[keV]           &  $8.05_{-0.06}^{+0.11}$    & $7.76_{-0.17}^{+0.18}$     & $8.26_{-0.13}^{+0.20}$      & $7.44_{-0.10}^{+0.11}$ & $8.39_{-0.76}^{+0.21}$ \\
&  $\sigma_{\rm abs1}$[keV]    & $ < $ 0.15                           &  $0.37_{-0.27}^{+0.26}$   & $0.95_{-0.33}^{+0.41}$      &  $0.27_{-0.11}^{+0.12}$  & $0.37_{-0.21}^{+0.54}$\\
&  EW$_{\rm abs1}$[eV]$^b$  &  $241_{-88}^{+99}$             & $294_{-175}^{+147}$        & $474_{-451}^{+422}$        & $231_{-0.20}^{+0.22}$  &  $285_{-127}^{+111}$\\
&  $E_{\rm abs2}$[keV]           &  $9.78_{-0.18}^{+0.19}$   & $10.68_{-3.91}^{+1.51}$  & $12.07_{-2.47}^{+1.39}$     & $10.78_{-4.38}^{+1.15}$ & $13.86_{-0.66}^{+1.56}$\\
&  $\sigma_{\rm abs2}$[keV]    & $0.43_{-0.17}^{+0.18}$     & $2.62_{-1.21}^{+1.51}$     & $2.91_{-1.49}^{+3.26}$      &  $2.82_{-1.52}^{+3.15}$ & $1.67_{-0.88}^{+1.24}$\\
&  EW$_{\rm abs2}$[eV]$^b$  &  $455_{-170}^{+187}$           & $963_{-960}^{+519}$        & $1577_{-1255}^{+516}$     & $1058_{-550}^{+387}$ & $476_{-318}^{+317}$ \\
&   $\chi^2/\nu$                        &  106.5/102                                                   & 106.4/113                             & 120.1/103                         & 163.3/141  & 65.9/68\\
&   $P(\chi^2/\nu)$                   & 0.36                                                                  & 0.66                                  & 0.12                                  & 0.10 & 0.55 \\

\enddata
\tablenotetext{a}{
Model 1 is a power-law with Galactic absorption and intrinsic
absorption (APL; XSPEC model wabs*zwabs*pow);
Model 2 is a power-law with Galactic absorption, intrinsic absorption, 
and a notch absorber (APL+No; XSPEC model
wabs*zwabs*notch*pow); 
Model 3 is a power-law with Galactic absorption, intrinsic absorption, 
and a notch absorber (IAPL+No; XSPEC model
wabs*absori*notch*pow); 
Model 4 is a power-law with Galactic absorption, 
intrinsic absorption, and an absorption edge 
(APL+Ed; XSPEC model wabs*zwabs*zedge*pow); 
Model 5 is a power-law with Galactic absorption, 
intrinsic ionized absorption, and an absorption edge 
(IAPL+Ed; XSPEC model wabs*absori*zedge*pow); 
Model 6 is a power-law with Galactic absorption, 
intrinsic absorption, and two absorption lines 
(APL+AL; XSPEC model wabs*zwabs*[pow+zgauss+zgauss]); 
Model 7 is a power-law with Galactic absorption, 
intrinsic ionized absorption, and two absorption lines 
(IAPL+2AL; XSPEC model wabs*absori*[pow+zgauss+zgauss]); 
All model fits include the Galactic absorption toward the source (Stark et al. 1992).
$N_{\rm H}$ is the intrinsic column density expressed in units of 10$^{22}$~atoms~cm$^{-2}$.
All fits have been performed on the combined spectrum of images A, B and C of \apm. }

\tablenotetext{b}{EW stands for equivalent width, which is defined
as $EW=\int \frac{F_c-F_E}{F_c}dE$, where $F_c$ is the continuous
flux and $F_E$ is the flux in the absorber.}

\tablenotetext{c}{All absorption-line parameters are calculated for the rest frame.}

\tablenotetext{d}{All errors are for 90\% confidence unless mentioned otherwise with all
parameters taken to be of interest except absolute normalization.}

\tablenotetext{e}{$P(\chi^2/{\nu})$ is the probability of exceeding $\chi^{2}$ for ${\nu}$ degrees of freedom
if the model is correct.}

\end{deluxetable}


\clearpage
\begin{deluxetable}
{cccccccccc} \tabletypesize{\small}
\tablecolumns{10} \tablewidth{0pt} \tablecaption{Results from 
spectral fits using XSTAR to the \chandra\ and \xmm\ observations of \apm.
\label{tab:xsta}}


\startdata
8........& $\Gamma$                                           &1.77$_{-0.04}^{+0.04}$& 1.94$_{-0.04}^{+0.04}$   & 2.07$_{-0.02}^{+0.03}$         & 2.20$_{-0.03}^{+0.03}$              & 1.93$_{-0.04}^{+0.04}$  \\
                   &   $N_{\rm H}$                            &4.5$_{-0.3}^{+0.5}$    & 5.5$_{-0.2}^{+0.7}$         & 5.3$_{-0.3}^{+0.2}$              & 5.2$_{-0.4}^{+0.2}$                & 10.5$_{-3.9}^{+1.1}$ \\
                    &   $N_{\rm H}$$_{\rm abs1}$     & 100                                &17$\pm$2                          & 39                                         & 12$_{-3}^{+2.0}$               & 8.5$_{-3.5}^{+3.8}$  \\
                    &   $\log\xi$$_{\rm abs1}$            & 4.0$_{-0.1}^{+0.08}$    &3.1$_{-0.7}^{+0.12}$        & 3.6$_{-0.3}^{+0.9}$              & 3.1$_{-0.17}^{+0.05}$                & 3.3$_{-0.6}^{+0.7}$      \\
                   &   $z_{\rm abs1}$                       & 3.2$\pm$0.10                &3.0$\pm$0.1                       & 3.1$\pm$0.1                          & 3.3$\pm$0.1                              & 2.94$\pm$0.10  \\
                  &   $N_{\rm H}$$_{\rm abs2}$     & 10$_{-5}^{+8}$                  & 14$_{-8}^{+4}$               & 83                                       & 10$_{-2}^{+2}$               & 40$_{-32}^{+40}$ \\
                  &   $\log\xi$$_{\rm abs2}$             & 3.2$_{-0.3}^{+0.2}$    &3.5$_{-0.4}^{+0.7}$          & 3.9$_{-0.3}^{+0.2}$              & 3.1$_{-0.05}^{+0.13}$                 & 3.9$_{-0.5}^{+0.5}$     \\
                  &   $z_{\rm abs2}$                          & 2.4$\pm$0.1               &2.0$\pm$0.1                       & 1.7$\pm$0.1                          & 2.3$\pm$0.1                             & 1.48$\pm$0.07 \\
               &   $\chi^2/\nu$                                & 114.3/103                  & 118.3/116                          & 129.2/104                              & 167.2/142                               & 71.5/69  \\
                  &   $P(\chi^2/\nu)$                            & 0.21                          & 0.42                                  & 0.05                                      & 0.07                                        & 0.40    \\
             &               &                       &             &                &  &    \\

9........& $\Gamma$                                       &  1.79$_{-0.04}^{+0.03}$             & 1.96$_{-0.10}^{+0.05}$                     & 2.08$_{-0.02}^{+0.03}$                       &2.32$_{-0.09}^{+0.1}$          & 1.93$_{-0.04}^{+0.03}$  \\
                   &   $N_{\rm H}$                      & 4.4$_{-0.5}^{+0.5}$                                 &5.7$_{-0.8}^{+0.3}$                           & 5.2$_{-0.23}^{+0.20}$                         &5.5$_{-0.3}^{+0.4}$               & 10.5$_{-1.2}^{+1.0}$  \\
                    &   $K_{\rm direct}$                 & $ < $ 1$\times$10$^{-5}$          &2.9$_{-2.2}^{+1.0}$$\times$10$^{-5}$&1.2$_{-1.2}^{+4.9}$$\times$10$^{-5}$  &1.4$_{-0.2}^{+0.2}$ $\times$10$^{-4}$ &   $ < $ 5$\times$10$^{-5}$   \\

                   &   $N_{\rm H}$$_{\rm abs1}$ &   100                                           & 50                                                      & 100                                                       & 72$_{-22}^{+30}$                             &  18$_{-12}^{+36}$ \\                                
                    &   $\log\xi$$_{\rm abs1}$        & 3.8$_{-0.3}^{+0.07}$                                 & 3.1$_{-0.07}^{+0.14}$                       & 3.3$_{-0.05}^{+0.06}$                          & 3.0$_{-0.07}^{+0.07}$           & 3.3$_{-0.7}^{+0.5}$     \\
                  &   $z_{\rm abs1}$                    & 3.2$\pm$0.1                                 & 3.0                                                     & 3.04$\pm$0.1                                        & 3.2 $\pm$0.1                             & 2.95 $\pm$0.1  \\
                 &   $K_{\rm abs1}$                   & 8$_{-1}^{+2}$$\times$10$^{-5}$   &7.8$_{-1.7}^{+1.6}$$\times$10$^{-5}$&5.8$_{-0.7}^{+0.6}$$\times$10$^{-5}$  &1.2$_{-0.2}^{+0.3}$$\times$10$^{-4}$   &1.1$_{-0.4}^{+3.2}$  $ \times$10$^{-4}$     \\

                  &   $N_{\rm H}$$_{\rm abs2}$& 22                                                  & 50$_{-7}^{+12}$                               &      100                                                  & 100                             & 100  \\
                   &   $\log\xi$$_{\rm abs2}$       & 3.0$_{-0.3}^{+0.4}$                                & 3.5$_{-0.6}^{+0.6}$                          & 4.0$_{-0.3}^{+0.08}$                           & 3.1$_{-0.2}^{+0.3}$               & 3.8$_{-0.9}^{+0.2}$     \\
                  &   $z_{\rm abs2}$                   & 2.3$\pm$0.1                                   & 2.0                                                    & 1.6$\pm$0.1                                          & 2.3$\pm$0.1                            & 1.5$\pm$0.1 \\
                  &   $K_{\rm abs2}$                  & 5$_{-1}^{+2}$$\times$10$^{-5}$  & 5$_{-1.2}^{+1.6}$  $\times$10$^{-5}$&1.2$_{-0.2}^{+0.2}$$\times$10$^{-4}$ & 0.8$_{-0.3}^{+0.4}$$\times$10$^{-4}$  &6.3$_{-0.3}^{+2.5}$$\times$10$^{-4}$     \\

                &   $\chi^2/\nu$                         & 116.9/101                                       & 116.2/114                                           & 116.2/102                                             & 156./140                       & 71.7/67 \\
                  &   $P(\chi^2/\nu)$                     & 0.13                                               & 0.43                                                   & 0.16                                                    & 0.17                              & 0.33    \\


\enddata

\tablenotetext{a}{ 
Model 8 is a power-law with Galactic absorption, 
intrinsic neutral absorption, and two intrinsic ionized absorbers 
(APL + 2IA; XSPEC model wabs*zwabs*warmabs*warmabs(pow)).
Model 9 is a power-law with Galactic absorption, 
intrinsic neutral absorption, and two partially covered intrinsic ionized absorbers
(APL + PC*(2IA); XSPEC model wabs*zwabs*(pow + warmabs*pow + warmabs*pow)).
In the APL + PC*(2IA) model 
$K_{\rm direct}$, $K_{\rm abs1}$ and $K_{\rm abs2}$ are the photon flux densities (photons~keV$^{-1}$~cm$^{-2}$~s$^{-1}$) at 1~keV of the
one non-absorbed and two absorbed components, respectively.  
All model fits include the Galactic absorption toward the source (Stark et al. 1992).
All fits have been performed on the combined spectrum of images A, B and C of \apm.
$N_{\rm H}$ is the intrinsic column density expressed in units of 10$^{22}$~atoms~cm$^{-2}$.
We note that in several cases (especially for spectral fits using model 9) the value of the column density of the ionized absorber 
reaches the maximum allowable value in our XSTAR model of 1 $\times$ 10$^{24}$~atoms~cm$^{-2}$.
The errors for several parameters are not listed because the fits in these cases could not provide any useful
constraints. All listed errors are at the 68\% confidence level.}

\end{deluxetable}

\clearpage
\begin{deluxetable}
{cccccccccccc} \tabletypesize{\scriptsize}
\tablecolumns{12} \tablewidth{0pt} \tablecaption{Projected maximum outflow velocities, mass-outflow rates
and efficiencies of outflows in \apm\ ${}^{a}$.
\label{tab:epk}}

\tablehead{
\colhead{Epoch} & \colhead{Instr.}  & \colhead{$v_{\rm
abs1}$} & \colhead{\lnh (abs1)} & \colhead{$\dot{M}$ (abs1)} &
\colhead{$\epsilon_K$ (abs1)} &  \colhead{$v_{\rm abs2}$} &
\colhead{\lnh (abs2)} &
\colhead{$\dot{M}$ (abs2)}  & \colhead{$\epsilon_K$ (abs2)}\\
& & [$c$] &  &  [$M_\odot \mu_L^{-1}{\rm yr}^{-1}$] &   & [$c$]
& & [$M_\odot \mu_L^{-1}{\rm yr}^{-1}$] &  }

\startdata

1 & ACIS S3  & $0.18_{-0.01}^{+0.01}$ & 22.9$\pm$0.4 & $393^{+403}_{-276}$ & $0.014^{+0.014}_{-0.01}$ & $0.36_{-0.02}^{+0.02}$  &  23.1$\pm$0.5 & $1229^{+1098}_{-792}$ & $0.17^{+0.15}_{-0.11}$\\

2 & EPIC pn  & $0.20_{-0.04}^{+0.03}$ & 23.1$\pm$0.3 & $656^{+846}_{-480}$ & $0.026^{+0.034}_{-0.019}$ & $0.46_{-0.06}^{+0.08}$ &  23.4$\pm$0.3 & $3141^{+899}_{-2246}$ & $0.68^{+0.20}_{-0.49}$\\

3 & EPIC pn  & $0.21_{-0.05}^{+0.05}$ & 23.2$\pm$0.3 & $891^{+1288}_{-651}$ & $0.04^{+0.06}_{-0.03}$ & $0.53_{-0.25}^{+0.07}$ &  23.6$\pm$0.3 & $5508^{+6437}_{-3973}$ & $1.61^{+1.89}_{-1.16}$\\

4 & EPIC pn  & $0.12_{-0.02}^{+0.02}$ & 22.8$\pm$0.3 &  $209^{+250}_{-158}$ & $0.003^{+0.004}_{-0.002}$ & $0.41_{-0.06}^{+0.06}$ &  23.4$\pm$0.4 & $2549^{+2593}_{-1719}$ & $0.46^{+0.47}_{-0.31}$\\

5 & ACIS S3  & $0.21_{-0.03}^{+0.03}$ & 23.0$\pm$0.7 & $609^{+542}_{-396}$ & $0.03^{+0.03}_{-0.02}$ & $0.61_{-0.04}^{+0.07}$ &  23.0$\pm$0.3 & $1593^{+1916}_{-1165}$ & $0.62^{+0.74}_{-0.45}$ \\




\enddata

\tablenotetext{a}{The estimated values of the outflow properties
were based on fits that assumed an absorbed power-law model with
two absorption lines. The values of $\dot{M}$ and $\epsilon_K$ are
obtained by equations 1 and 2, respectively,
assuming $M_{\rm BH}\sim 10^{12}\mu_L^{-1}M_\odot$ (see \S 3.2) and $L_{\rm
bol}=7\times10^{15}\mu_L^{-1}L_\odot$ (Irwin et al. 1998; Reichers et al. 2009).}

\end{deluxetable}

\clearpage
\begin{deluxetable}
{ccccccccccc} \tabletypesize{\small}
\tablecolumns{11} \tablewidth{0pt} \tablecaption{
The minimum and maximum
energies and velocities of the high-energy absorption features in \apm.
\label{tab:vma}}

\tablehead{
 \colhead{Epoch} & \colhead{Instrument} & &
\colhead{$E_{\rm min}$} & \colhead{$E_{\rm max}$} &
\colhead{$v_{\rm min}$} & \colhead{$v_{\rm max}$} \\
& & & [keV] & [keV] & [$c$] & [$c$]
}

\startdata

 1 & ACIS S3 & & 8.05$\pm$0.17 & 10.64$\pm$0.27 & 0.18$\pm$0.02 & 0.43$\pm$0.02 \\

 2  & EPIC pn & &7.54$\pm$0.74  & 15.04$\pm$1.22 & 0.12$\pm$0.09 & 0.67$\pm$0.04 \\

 3 & EPIC pn & & 6.42$\pm$0.82  & 17.94$\pm$2.96 & 0.04$\pm$0.04 & 0.76$\pm$0.07 \\

 4 & EPIC pn & & 6.94$\pm$0.35  & 16.31$\pm$1.17 & 0.04$\pm$0.04 & 0.71$\pm$0.04 \\

 5 & ACIS S3  & & 7.36$\pm$0.56 & 15.35$\pm$2.12 & 0.09$\pm$0.07 & 0.68$\pm$0.07 \\


\enddata

\tablenotetext{a}{$E_{\rm min}$ and $E_{\rm max}$ are estimated from spectral fits using model 6 of Table 2.}


\end{deluxetable}

\clearpage
\begin{figure*}
\centerline{\includegraphics[width=16cm]{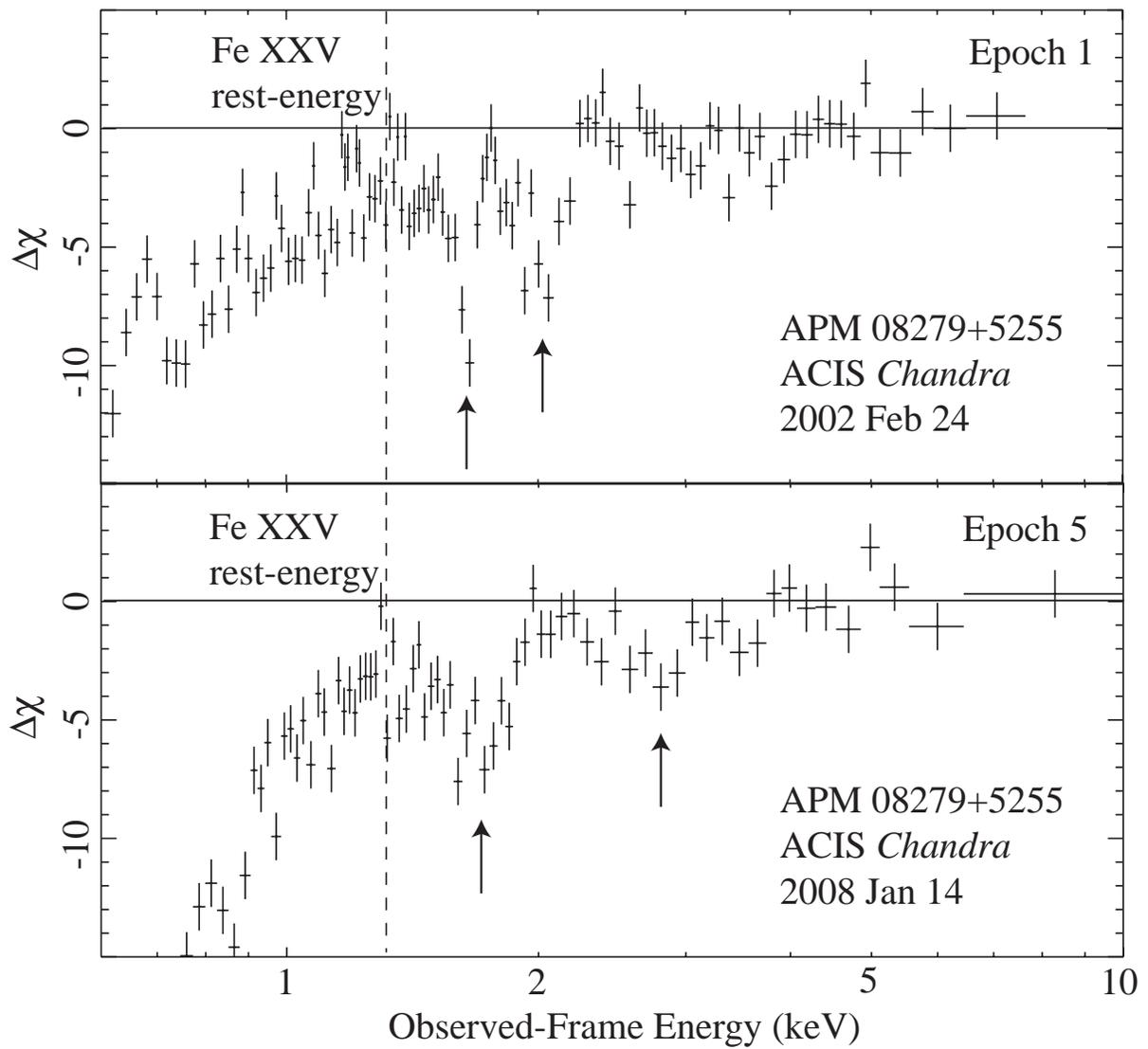}}
\caption{ \small $\Delta\chi$ residuals between the best-fit 
Galactic absorption and power-law model and the {\sl Chandra} ACIS spectra of \apm. 
This model is fit to events with energies lying within the ranges 4.5--10~keV.
The arrows indicate the best-fit energies of the absorption
lines of the first and second outflow components for epoch 1 (top panel) and epoch 5 (lower panel) obtained in fits that used model 6 of Table~2. 
\label{fig1.eps}}
\end{figure*}

\clearpage
\begin{figure*}
\centerline{\includegraphics[width=16cm]{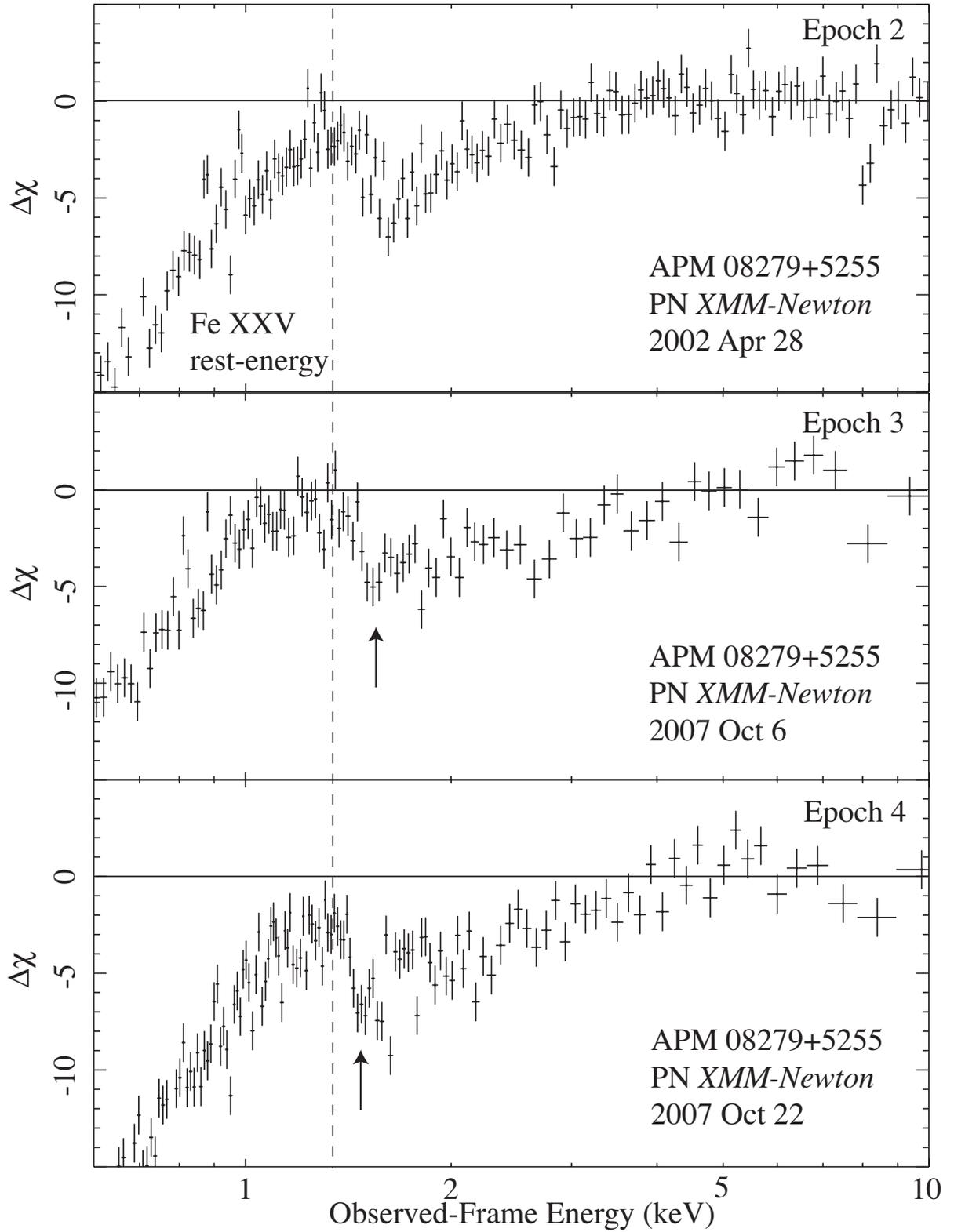}}
\caption{ \small $\Delta\chi$ residuals between the best-fit 
Galactic absorption and power-law model and the \xmm\ pn spectra of \apm. 
This model is fit to events with energies
lying within the range 4.5--10~keV.
The arrows indicate the best-fit energies of the absorption
line of the first outflow component for epoch 3 (middle panel) and epoch 4 (bottom panel) obtained in fits that used model 6 of Table~2. 
\label{fig2.eps}}
\end{figure*}

\clearpage
\begin{figure*}
\centerline{\includegraphics[width=16cm]{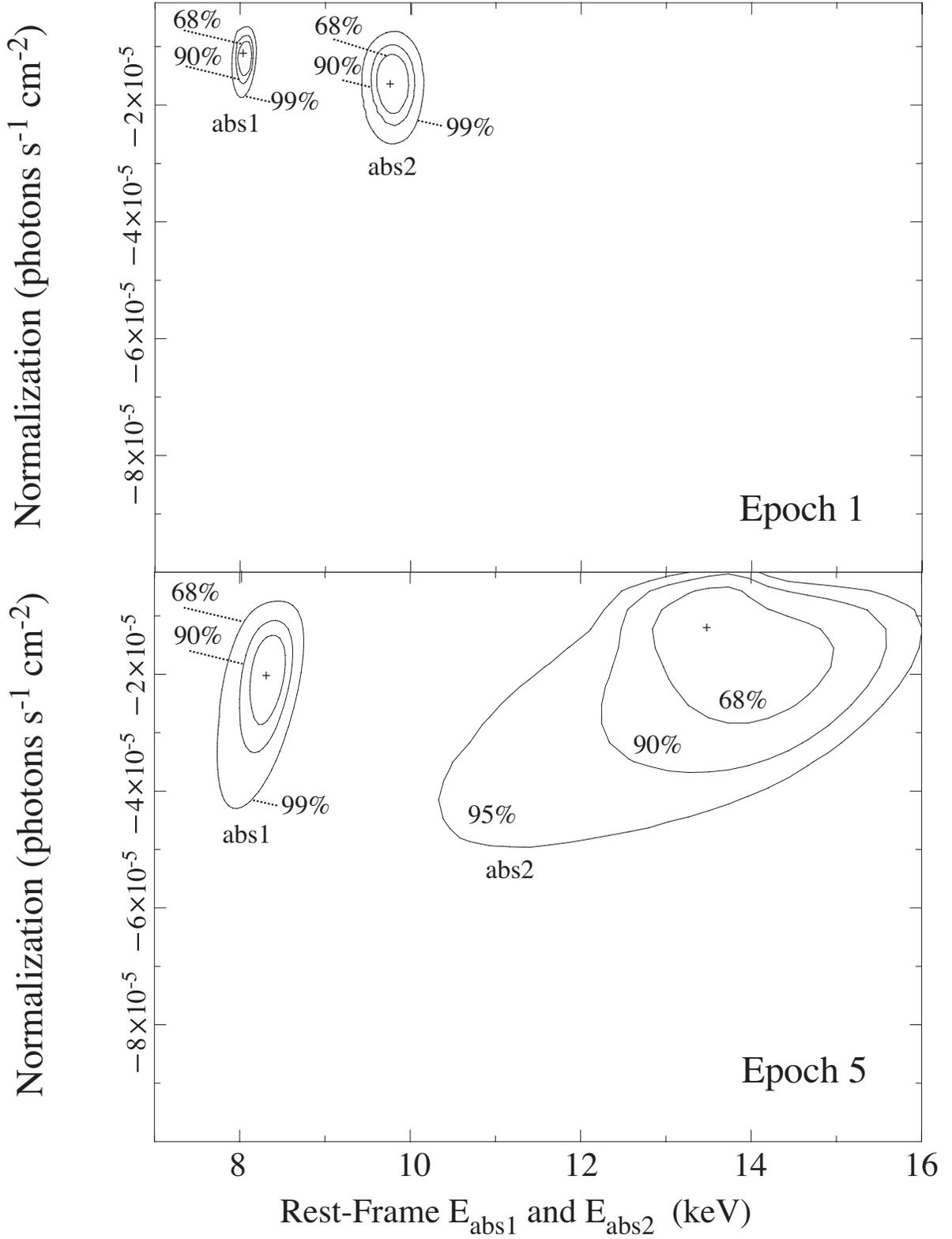}}
\caption{ \small 
68\%, 90\%, and 99\%  confidence contours between the normalizations 
of the absorption lines at $E_{\rm abs1}$ and $E_{\rm abs2}$
and the respective energies $E_{\rm abs1}$ and $E_{\rm abs2}$ for the \chandra\ 2002 (top panel) and \chandra\ 2008 (bottom panel) 
observations of \apm\ assuming fit 6 of Table 2. The 99\% confidence contour of component $abs2$ for epoch 5 is not well constrained and 
the 95\% contour is shown instead.
\label{fig3.eps}}
\end{figure*}

\clearpage
\begin{figure*}
\centerline{\includegraphics[width=16cm]{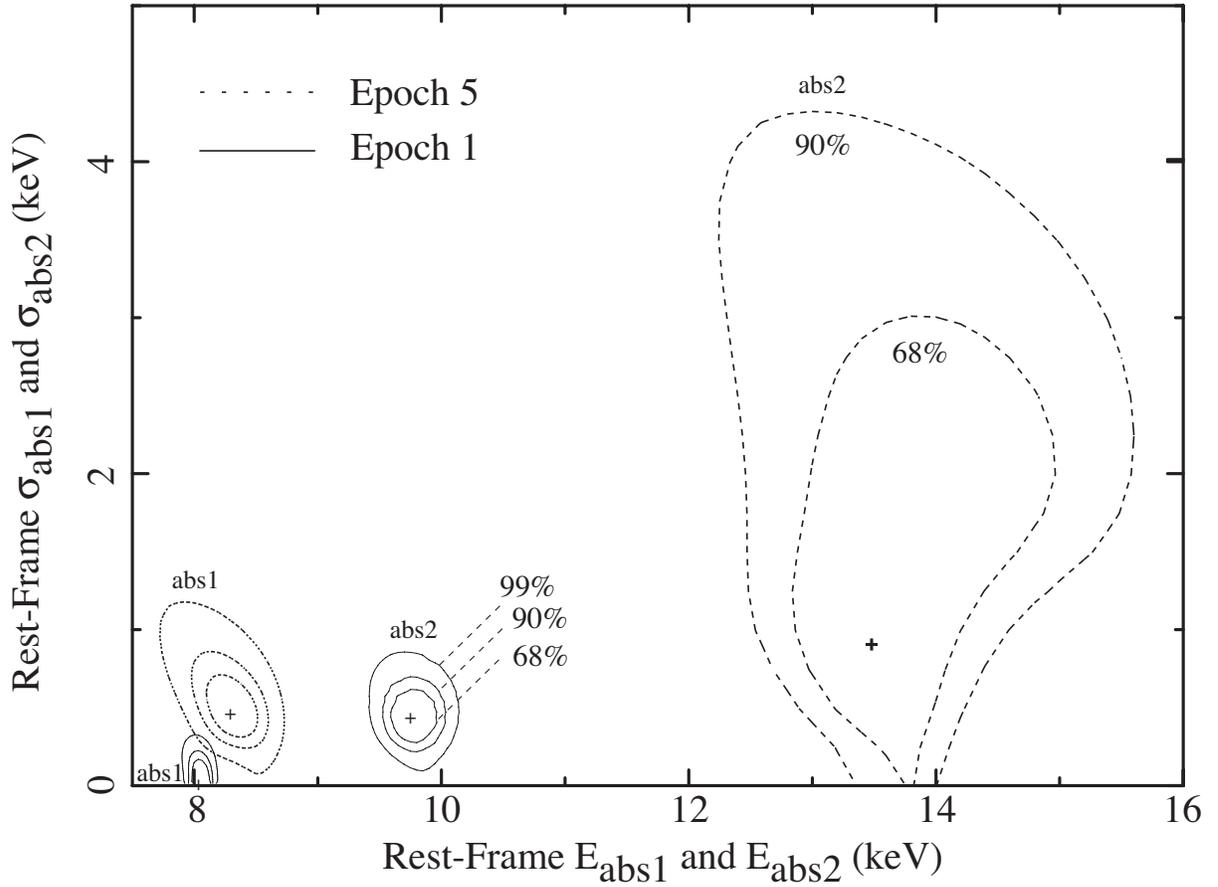}}
\caption{ \small 
68\%, 90\% and 99\% confidence contours between best-fit energies $E_{\rm abs1}$ and $E_{\rm abs2}$
and energy widths $\sigma_{\rm abs1}$ and $\sigma_{\rm abs2}$ of the absorption lines for the \chandra\ 2002 (Epoch 1)  and \chandra\ 2008 (Epoch 5) 
observations of \apm\ assuming fit 6 of Table 2. The 99\% confidence contour of component $abs2$ for epoch 5 is not well constrained 
and not shown in this plot.
\label{fig4.eps}}
\end{figure*}

\clearpage
\begin{figure*}
\centerline{\includegraphics[width=16cm]{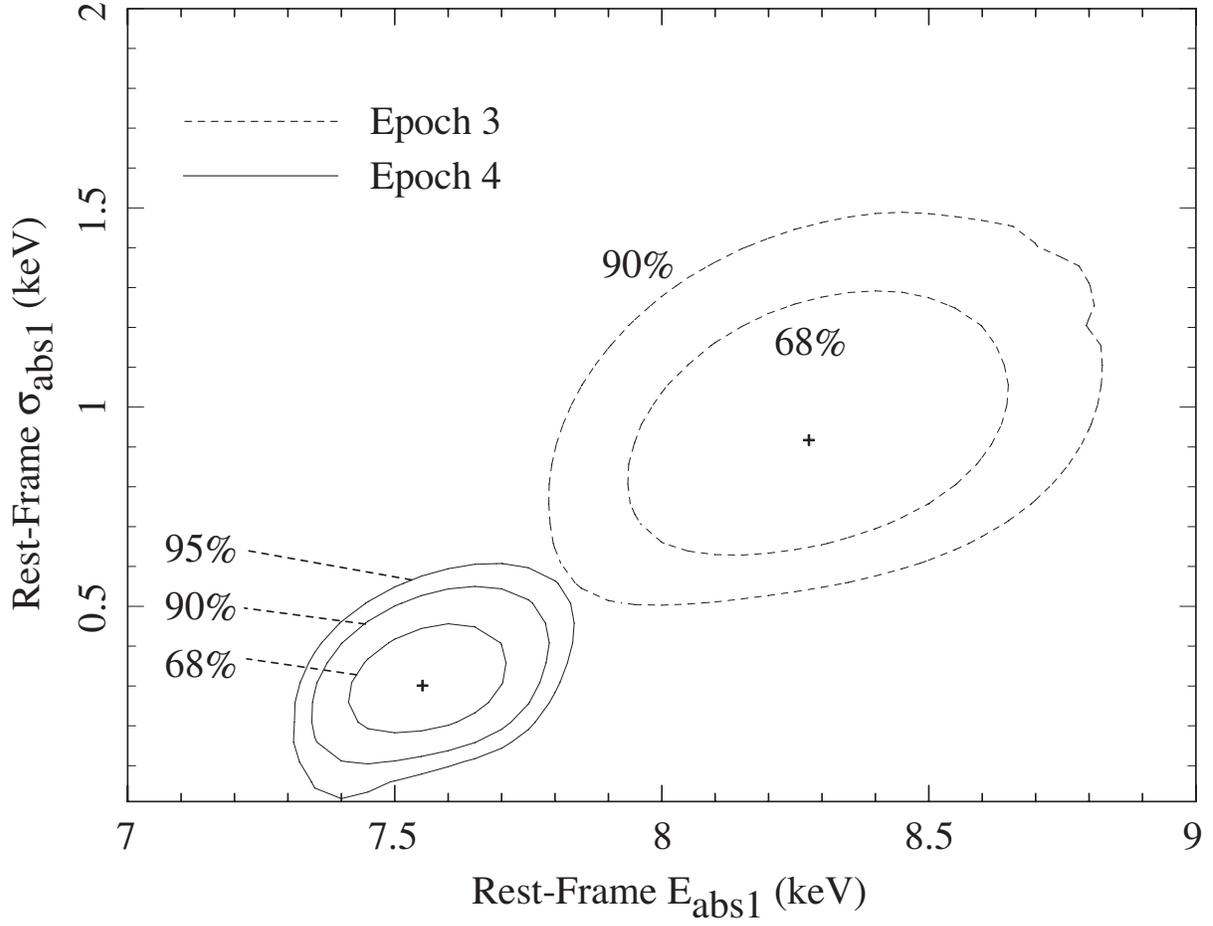}}
\caption{ \small 
68\%, 90\% and 95\% $\chi^{2}$ confidence contours of
$E_{\rm abs}$ versus $\sigma_{\rm abs}$ of the first absorption line in epoch 3 (dotted line)  
and epoch 4 (solid line) assuming the APL + 2 AL model (see model 6 of Table 2).
The 95\% confidence contour for epoch 3 is not well constrained 
and not shown in this plot.
\label{fig5.eps}}
\end{figure*}

\clearpage
\begin{figure*}
\centerline{\includegraphics[width=16cm]{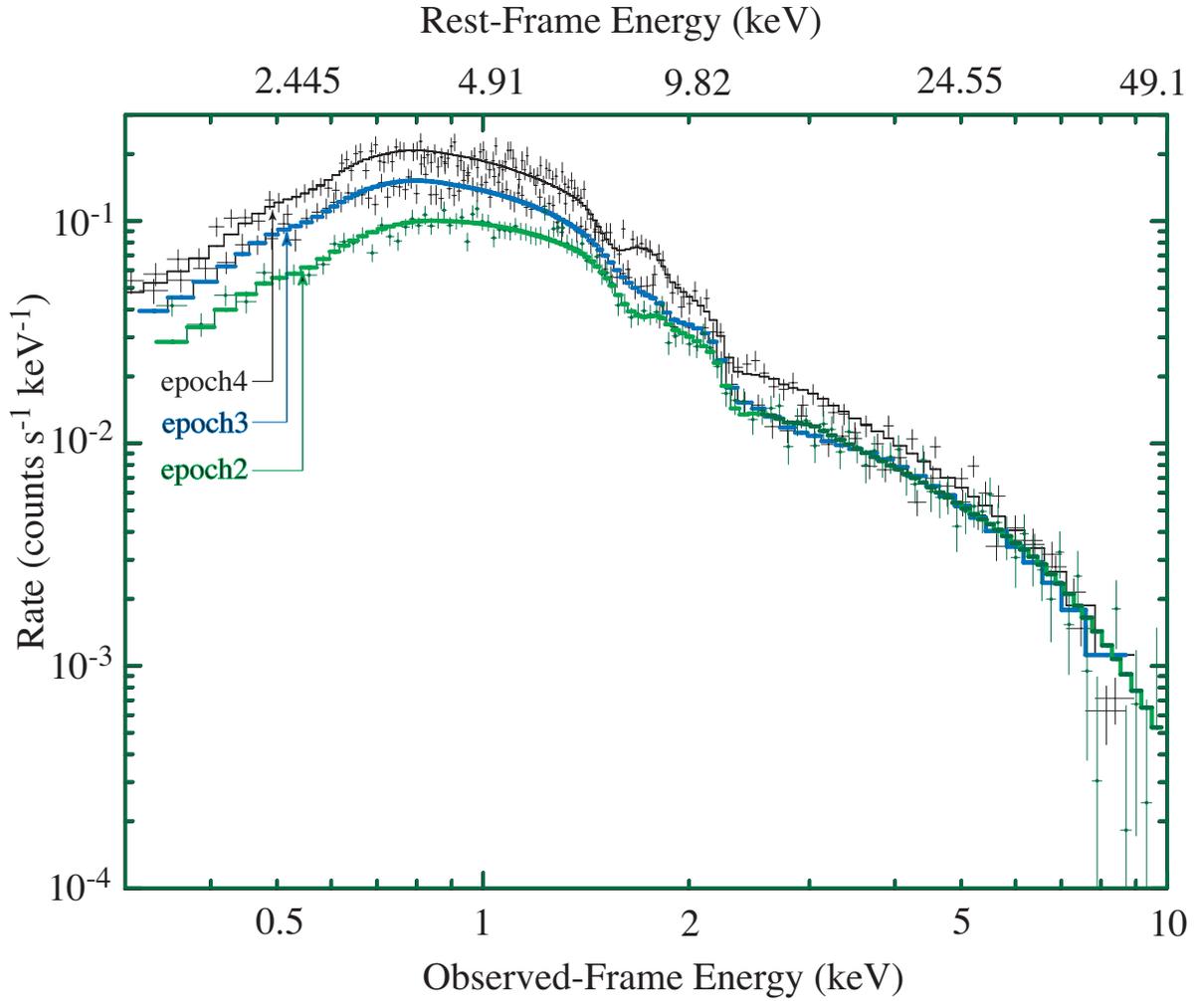}}
\caption{ \small The spectra and best-fit models (model 6 of Table 2) of the \xmm\
observations of \apm. We note the significant variability of the X-ray BALs (extending between $\approx$ 1.5--3.6~keV)
and continuum between epochs 3 and 4 that are separated by only $\sim$ 3.3~days (proper-time).
\label{fig6.eps}}
\end{figure*}

\clearpage
\begin{figure*}
\centerline{\includegraphics[width=16cm]{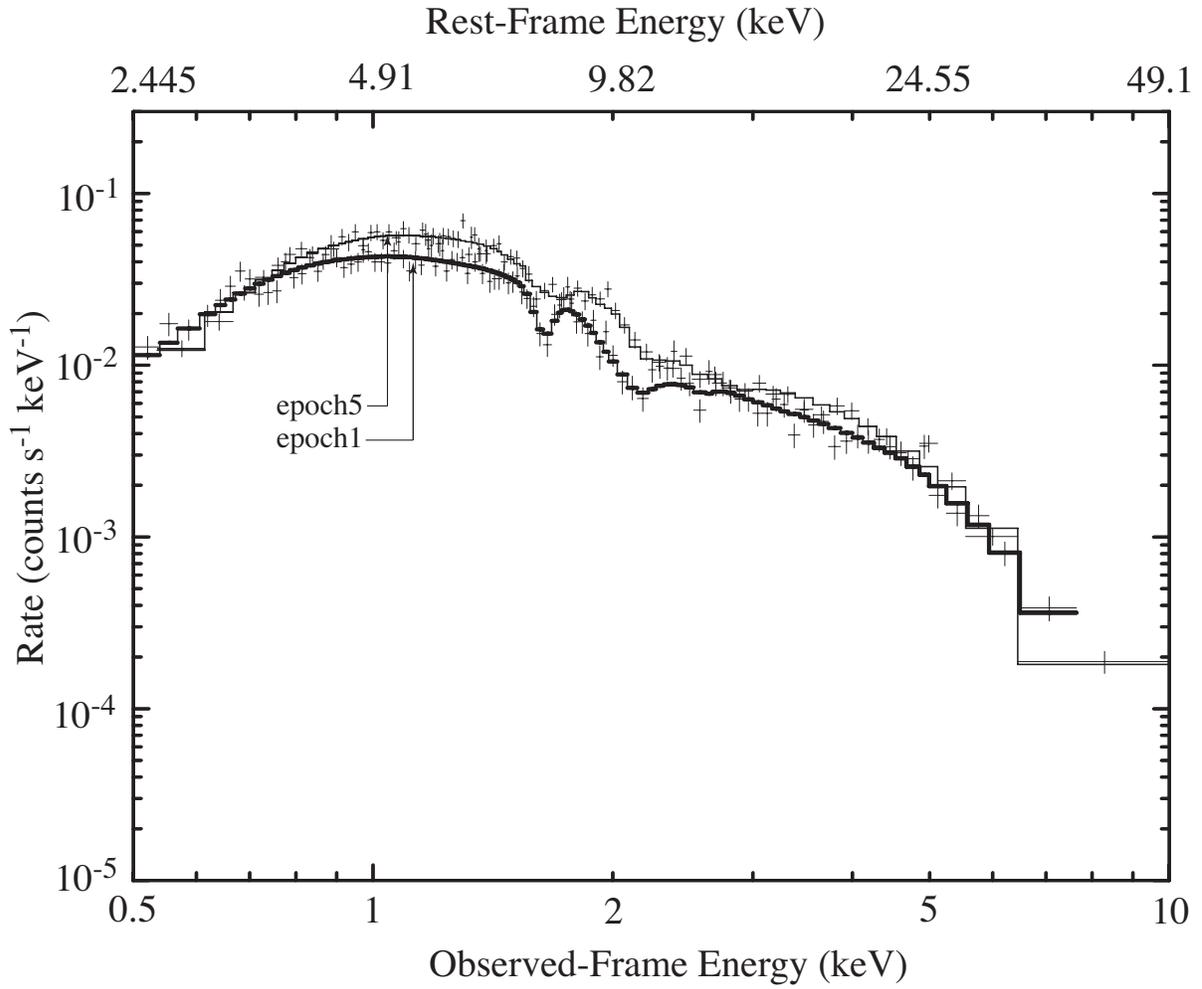}}
\caption{ \small The spectra and best-fit models (model 6 of Table 2) of the \chandra\ 
observations of \apm. We note the significant variability of the X-ray BALs extending between $\approx$ 1.5--3.6~keV.
\label{fig7.eps}}
\end{figure*}

\clearpage
\begin{figure*}
\centerline{\includegraphics[width=14cm]{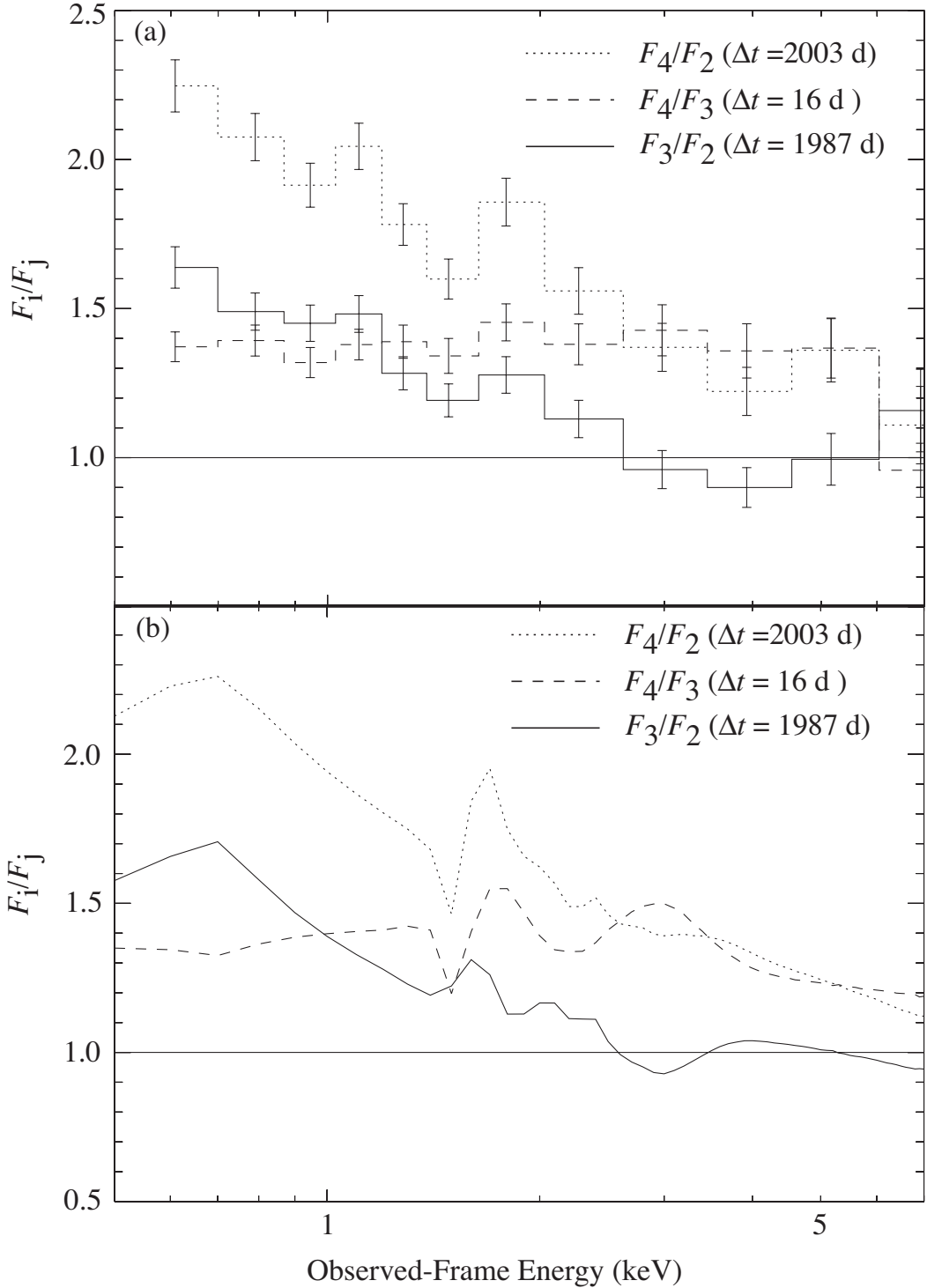}}
\caption{ \small {The ratios of the (a) observed and (b) best-fit-modeled (using model 6 of Table 2) flux densities
$F_{\rm i}/F_{\rm j}$ between the three \xmm\ observations of \apm\ listed in Table 1, where $i$ and $j$ represent the 
epochs compared, ${\Delta}t$ is the observed-frame time between epochs 
and $F$ is the flux density in units of counts~s$^{-1}$~keV$^{-1}$.
Note that a significant change of flux density is observed
between all observations over the entire spectrum of \apm.
The spectral change between epochs 3 and 4 is particularly remarkable
since it occurs over a period of only $\sim$~3.3~days (proper-time)
implying a size-scale of $\sim$~7.4 $\times$ 10$^{15}$~cm which is comparable 
to $r_{\rm ISCO}$ = 4.5 $\times$ 10$^{15}$~cm for the case of a Schwarzschild black hole in \apm.}
\label{fig8.eps}}
\end{figure*}

\clearpage
\begin{figure*}
\centerline{\includegraphics[width=14cm]{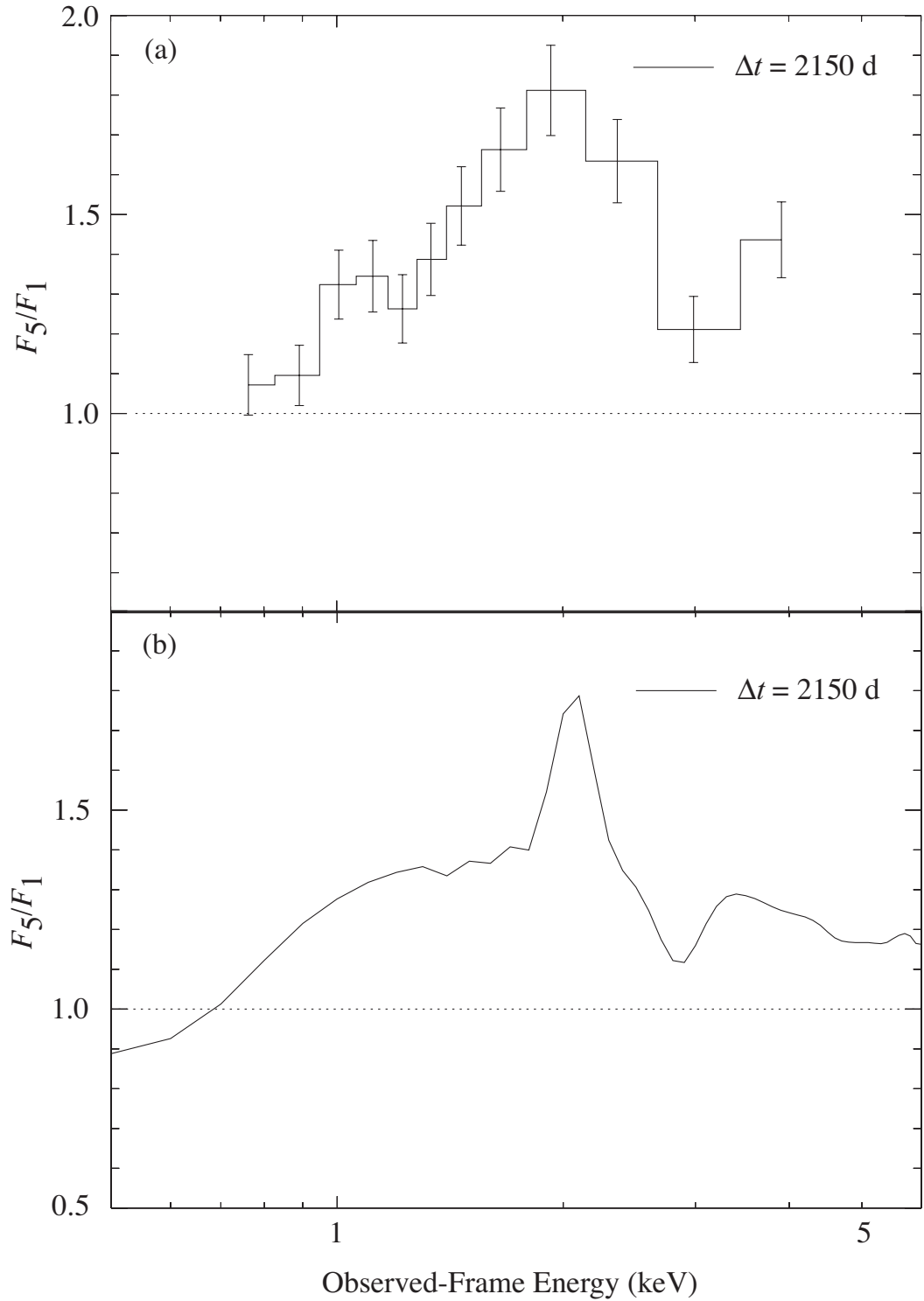}}
\caption{ \small 
{The ratios of the (a) observed and (b) best-fit-modeled (using model 6 of Table 2) flux densities
$F_{\rm 5}/F_{\rm 1}$ between the two {\sl Chandra} observations of \apm\ listed in Table 1, where 
$F_{1}$ and $F_{5}$ are the flux densities for epochs 1 and 5 respectively.
$\Delta$$t$ is the observed-frame time between epochs.
Note that a significant change of flux density is observed
near the region of the Fe BALs ($\sim$ 1.5--3.6~keV).}
\label{fig9.eps}}
\end{figure*}

\clearpage

\begin{figure*}
\centerline{\includegraphics[width=14cm]{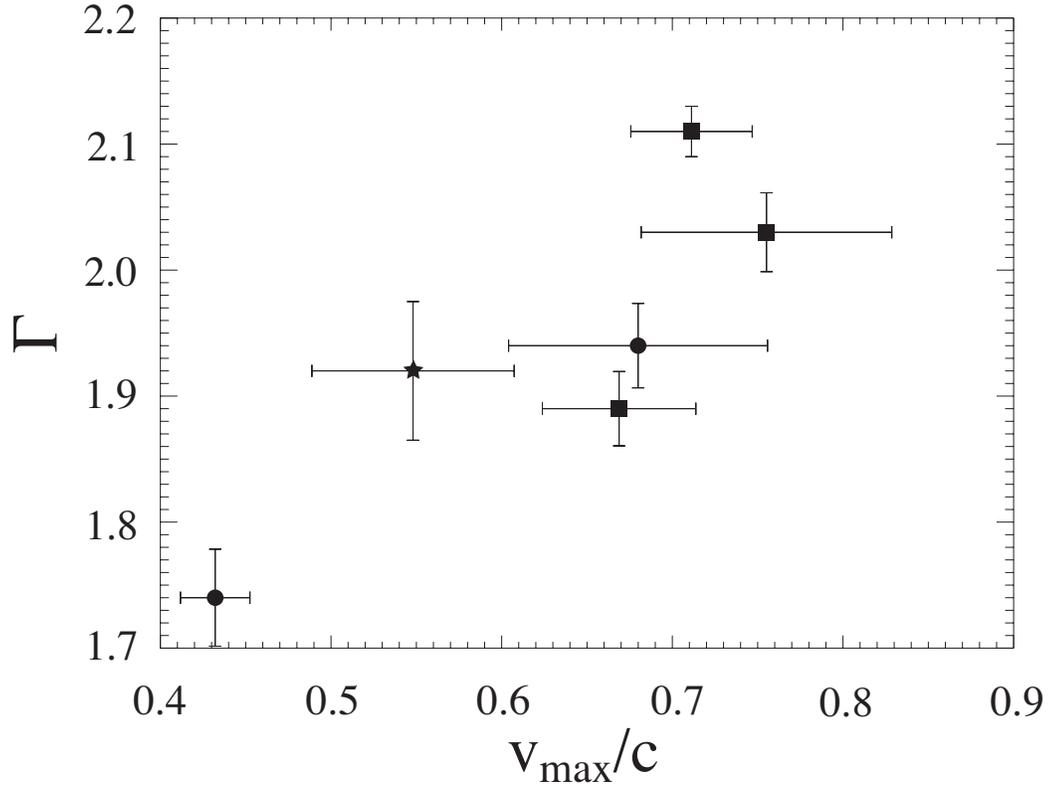}}
\caption{Maximum outflow velocity, $v_{\rm max}$, observed from \apm\ as a function of the X-ray photon index, ${\Gamma}$.
$v_{\rm max}$ and ${\Gamma}$ were derived from fits to the spectra of \apm\ with a model that included an absorbed power-law and two Gaussian absorption lines (model 6 of Table 2).  Errors shown are at the 68\% confidence level. Data shown with circles, a star, and squares are obtained from {\sl Chandra}, {\sl Suzaku} and \xmm\ observations respectively.
\label{fig10.eps}}
\end{figure*}

\clearpage

\begin{figure*}
   \centerline{\includegraphics[width=16cm]{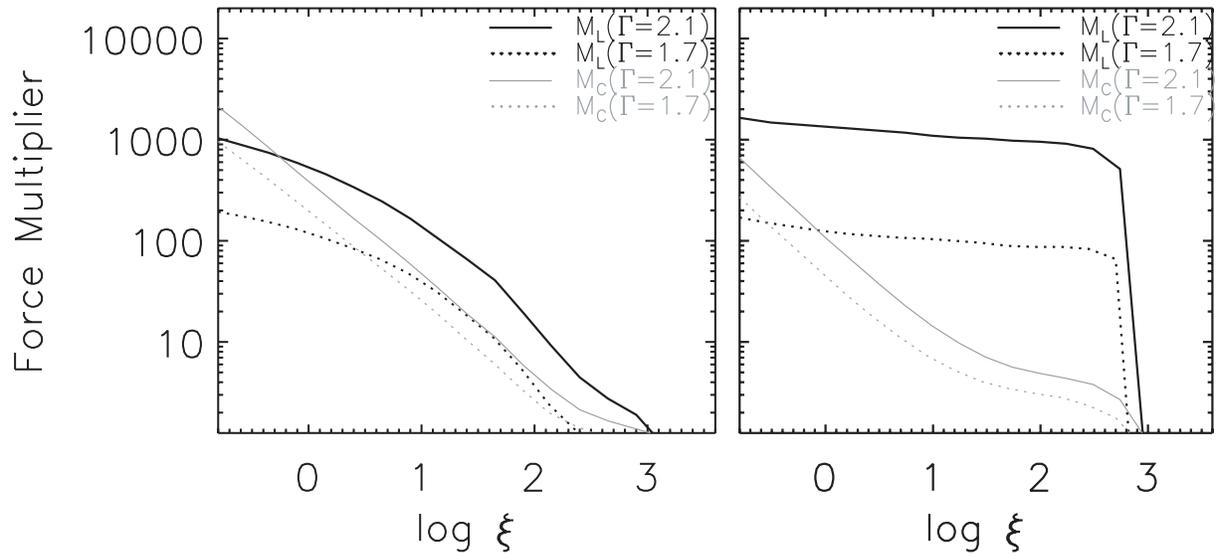}}
        \centering
      \caption{ The bound-free, $M_{\rm C}$, and bound-bound, $M_{\rm L}$, components of the force multiplier
are shown as a function of the ionization parameter. Force multipliers are calculated for 
SEDs with photon indices of $\Gamma$ = 2.1 (solid lines) and $\Gamma$ = 1.7 (dotted lines).
In the left panel we have assumed no absorbing shield, whereas, in the right panel 
the soft and hard SEDs have been attenuated by an absorber with $\lnh =22.8$.}
        \label{fig11.eps}
\end{figure*}

\clearpage
\begin{figure*}
\centerline{\includegraphics[width=16cm]{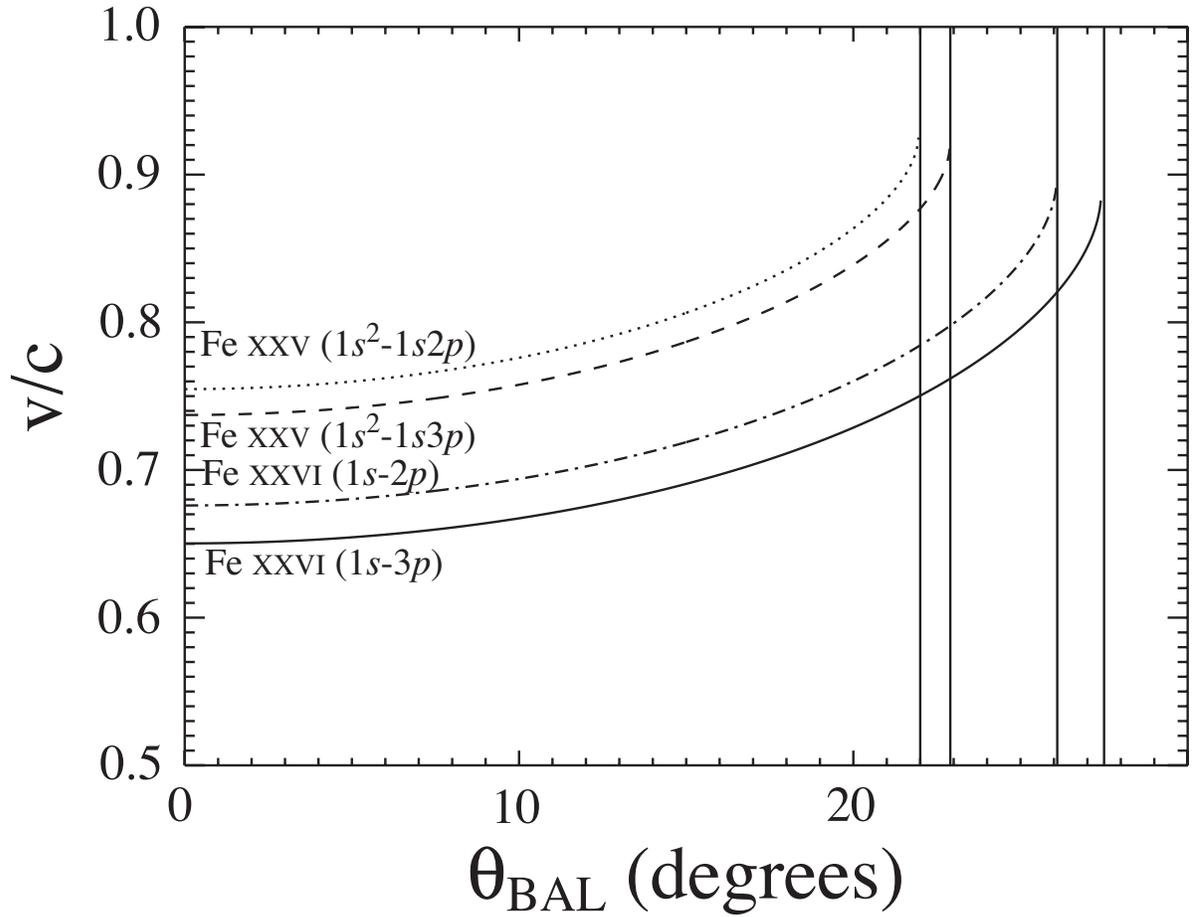}}
\caption{Maximum outflow velocity as a function of the angle between the outflow direction and our line of sight. 
Shown are the outflow velocities for the most likely resonance transitions responsible for the observed X-ray BALs in APM 08279+5255. 
The outflow velocities were calculated using the relativistic Doppler formula 
for the case of the maximum rest-frame energy of the X-ray BALs of $E_{max}$ = 17.9~keV (see Table ~5)
observed in epoch 3.
\label{fig12.eps}}
\end{figure*}

\end{document}